\documentclass[prd,aps,showpacs,tightenlines,floats,nofootinbib,preprintnumbers]{revtex4}

\usepackage{graphicx}
\usepackage{amsmath,amssymb,color}
\usepackage{hyperref}

\newcommand{\pfrac}[2]{\frac{\partial#1}{\partial#2}}
\newcommand{\ppfrac}[2]{\frac{\partial^2#1}{\partial#2^2}}
\newcommand{\ppmfrac}[3]{\frac{\partial^2#1}{\partial#2\partial#3}}
\newcommand{\xx}{\mathbf{x}}
\newcommand{\bb}{\mathbf{b}}

\definecolor{darkgreen}{cmyk}{0.85,0.2,1.00,0.2}

\begin{document}

\title{Equivalence principle violation in Vainshtein screened two-body systems}
\author{
Takashi Hiramatsu$^1$,
Wayne Hu$^2$,
Kazuya Koyama$^3$ and
Fabian Schmidt$^4$
}
\affiliation{
$^1$Yukawa Institute for Theoretical Physics, Kyoto
University, Kyoto 606-8502, Japan\\
$^2$Kavli Institute for Cosmological Physics, Department of Astronomy \&
Astrophysics, University of Chicago, Chicago, Illinois 60637, USA\\
$^3$Institute of Cosmology \& Gravitation, University of Portsmouth,
Dennis Sciama Building, Portsmouth, PO1 3FX, United Kingdom\\
$^4$California Institute of Technology, Mail Code 350-17, Pasadena, California
91125, USA
}

\preprint{YITP-12-78}

\begin{abstract}
In massive gravity, galileon, and braneworld explanations of cosmic
 acceleration, force modifications are screened by nonlinear  derivative
 self-interactions of the scalar field mediating that force.
 Interactions between the field of a central body (``$A$") and an
 orbiting body (``$B$") imply that body $B$ does not move as a test body
 in the field of body $A$ if the orbit is smaller than the Vainshtein
 radius of body $B$.    We find through numerical solutions of the joint
 field at the position of $B$ that the $A$-field Laplacian is nearly
 perfectly screened by the $B$ self-field,  whereas first derivative or
 net forces are reduced in a manner that scales with the mass ratio of
 the bodies as  $(M_B/M_A)^{3/5}$.    The latter causes mass-dependent
 reductions in the universal perihelion precession rate due to the fifth
 force, with deviations for the Earth-Moon system at the $\sim 4\%$ level.
In spite of universal coupling, which preserves the microscopic equivalence principle,  the
motion of macroscopic screened bodies depends on their mass providing in principle a means for testing the Vainshtein mechanism.
\end{abstract}

\maketitle

\section{Introduction}

The current acceleration of the cosmic expansion is one of
the most puzzling aspects of modern cosmology.
Aside from a cosmological constant whose smallness remains unexplained, the simplest
models typically involve an additional scalar field either implicitly or
explicitly.  Universal coupling of this field to matter would produce
gravitational strength fifth forces and naively be excluded by laboratory and
solar system bounds. Viable models must therefore implement a so-called
screening mechanism to hide fifth forces locally.

Screening mechanisms invoke nonlinearity in the  field equations to
change the nature of the fifth force in high density regions. For
example, the chameleon mechanism increases the mass of the field in
deep gravitational potentials \cite{Khoury:2003rn} whereas the
symmetron mechanism changes its coupling to matter \cite{Hinterbichler:2010es}.
A third possibility is the Vainshtein mechanism \cite{Vainshtein:1972sx,
Babichev:2010jd}, first introduced in the context of massive gravity to
suppress the propagation of additional helicity modes
\cite{vanDam:1970vg,Zakharov:1970cc}. Here nonlinear derivative
interactions of the field act to screen the fifth force within the 
so-called Vainshtein radius around a matter source. The Vainshtein
mechanism occurs not only in modern incarnations of Boulware-Deser
\cite{Boulware:1972zf} ghost-free massive gravity
\cite{deRham:2010kj, Chkareuli:2011te, Koyama:2011yg, Sbisa:2012zk} but also in Galileon cosmology
\cite{Nicolis:2008in, Deffayet:2009wt, Burrage:2010rs, Kaloper:2011qc,
DeFelice:2011th, Kimura:2011dc, deRham:2012fw} and braneworld
models. Indeed it is in the braneworld model of  Dvali, Gabadadze and
Porrati (DGP) \cite{Dvali:2000hr} that it has been best studied
\cite{Deffayet:2001uk, Lue:2002sw, Lue:2004rj, Koyama:2007ih,
Schmidt:2009sg,Schmidt:2009sv,Schmidt:2009yj}. 

Interestingly, these mechanisms are distinguished by how
screened bodies fall in external fields \cite{Hui:2009kc}.   As a
consequence of universal coupling, all unscreened test bodies fall
in the same way and obey a microscopic equivalence principle. In the
chameleon and symmetron models, screened bodies do not respond to
external fields.  In the Vainshtein mechanism they do, but only if
those fields have wavelengths long compared to the Vainshtein radius.
These differences arise because the self-field of the screened body and
the external field do not in general superimpose but rather interfere in
a manner dependent on the nonlinear interaction \cite{Hu:2009ua}.

In this paper, we consider the Vainshtein mechanism in the near-field limit.
In particular we study the motion of two bodies that are separated by much less
than their individual Vainshtein radii and look for apparent violations of the equivalence
principle. The two-body problem is particularly relevant since it has been shown that
for the orbit of test bodies, there is a universal anomalous precession rate
induced by a Vainshtein-screened scalar field that is
potentially measurable in next generation solar system tests of general relativity
\cite{Lue:2002sw,Dvali:2002vf}.   However
in the Earth-Moon system the Moon is screened on scales out to nearly a
parsec  for cosmologically motivated models and cannot be considered as
a test body in the Earth's field. In principle, this nonlinearity can affect the interpretation of lunar
ranging tests for anomalous precession and more generally lead to
results that depend on the nature of the orbiting body. 
To understand this system, the field must be solved jointly
in the presence of both sources. For definiteness, we will implement the
Vainshtein mechanism in the DGP model but our results readily apply to the Galileon and massive
gravity incarnations as well.

The remainder of the paper is organized as follows.
We briefly review the Vainshtein mechanism and the
spherically symmetric one-body solution in Sec.~\ref{sec:DGP}.
In Sec.~\ref{sec:superposition}, we discuss violation of the
superposition principle in the two-body case. We present numerical
results and their scaling with the two-body parameters in
Sec.~\ref{sec:results}.  Details of the numerical scheme is given in the
Appendix.  We discuss the implications of these results
in  Sec.~\ref{sec:conclusion}.  

\section{Vainshtein Mechanism}
\label{sec:DGP}

\subsection{DGP example}

As an example of models that accommodate the Vainshtein mechanism, we
consider the DGP  braneworld model \cite{Dvali:2000hr}. In the DGP
model, there is a dynamical degree of freedom representing the bending
of the brane embedded in the five-dimensional bulk that we denote $\phi(\mathbf{x}, t)$.
In the quasistatic limit where its time derivatives can be neglected, its equation of
motion becomes \cite{Koyama:2007ih,Lue:2002sw}
%
\begin{equation}
 3\beta(t)\frac{\nabla^2}{a^2}\phi +
  \frac{1}{a^4} N[\phi,\phi] = 8\pi G \delta\rho, \label{eq:phievo}
\end{equation}
%
where
%
\begin{align}
  \beta(t) &= 1\pm 2Hr_c\left(1+\frac{\dot{H}}{3H^2}\right), \\
 N[\phi_A,\phi_B] &= r_c^2( \nabla^2\phi_A\nabla^2\phi_B
    - \nabla_i\nabla_j\phi_A\nabla^i\nabla^j\phi_B ) ,\label{eq:defN}
\end{align}
and $\delta \rho= \rho-\bar \rho$, the density fluctuation from the cosmic mean.
%
Spatial derivatives here are in comoving coordinates.
The key parameter in this model is the crossover scale 
%
\begin{equation}
 r_c = \frac{G_5}{2 G},
\end{equation}
%
the ratio between the five-dimensional Newton constant $G_5$ and the
four-dimensional one $G$.
In the main part of this paper, we will consider a binary system such as Earth and Moon.
We expect that the quasistatic limit is valid even in such a system.
Considering the dynamics of this system, the typical time scale of the system is
determined from the velocity of the Kepler motion, $\mathbf{v}_{\rm Kepler}$.
Hence the time dependence of $\phi$ would be 
$c^{-1}\dot{\phi} \sim c^{-1}\mathbf{v}_{\rm Kepler}\cdot\nabla\phi$, and thus the
time dependence of the scalar field is suppressed by 
$|\mathbf{v}_{\rm Kepler}|/c \sim 10^{-6}$ in the Earth-Moon system, which
validates the quasistatic limit.

The $+$ sign in $\beta(t)$ corresponds to the normal branch solution
while the $-$ sign corresponds to the self-accelerating solution.
In this paper, we only consider the solutions with $\beta>0$ since
$\beta<0$ is associated with ghost
instabilities \cite{Koyama:2007za}.
Furthermore, since we are interested in static solutions at the current
epoch $a=1$, we set  $\beta=1$.  All results can be rescaled to the 
general $\beta$ case by replacing $r_c$ with $r_c/\sqrt{\beta}$ and densities
$\delta \rho$ with $\delta \rho/\beta$, or equivalently the masses of all bodies.

We have written the nonlinear operator  $N[\phi_A,\phi_B]$ in bilinear
form allowing for two separate fields since in the two body calculation
that follows it will be useful to consider the interference between the two
individual fields. It is the nonlinearity of this
operator that is responsible for both the Vainshtein mechanism and the
lack of a superposition principle for the brane bending mode.

\subsection{One-body solution}
\label{sec:spherical}

The Vainshtein mechanism and the scale associated with it can be
illustrated with simple analytic one-body solutions.  For
a spherically symmetric object with a
top-hat constant density, the scalar field equation (\ref{eq:phievo}) reduces to
%
\begin{equation}
  3\left(\frac{d^2\phi}{dR^2} + \frac{2}{R}\frac{d\phi}{dR}\right)
  + r_c^2\left[\frac{2}{R^2}\left(\frac{d\phi}{dR}\right)^2
  + \frac{4}{R}\frac{d^2\phi}{dR^2}\frac{d\phi}{dR}\right]
  = 8 \pi G \delta\rho,\label{eq:phisph}
\end{equation}
%
where the top-hat density profile is
%
\begin{equation}
  \delta\rho(R) =
\begin{cases}
  \delta\rho_0 & R\leq r_s \\
  0   & R> r_s
\end{cases},\label{eq:tophat}
\end{equation}
%
with $r_s$ as the radius of the source. 
Equation~(\ref{eq:phisph}) can be integrated by multiplying both
sides by $R^2 dR$ , resulting in
%
\begin{equation}
 3 R^2\frac{d\phi}{dR}
     + 2 R r_c^2\left(\frac{d\phi}{dR}\right)^2
 =8\pi G \int_0^R\!\delta\rho(R')R'^2\,dR' =
\begin{cases}
  8 \pi G \delta\rho_0 R^3/3 & R\leq r_s, \\
  8 \pi G \delta\rho_0r_s^3/3   & R>r_s.
\end{cases}
\end{equation}
%
As this is a quadratic equation for $d\phi/dR$ we immediately obtain
%
\begin{equation}
  \frac{d\phi}{dR} = \frac{3 R}{4r_c^2}\times
\begin{cases}
g(r_s) & R\leq r_s,\\
g(R) & R>r_s,
\end{cases}\label{eq:phisph2}
\end{equation}
%
where
%
\begin{equation}
g(R) = \sqrt{1+\left(\frac{r_*}{R}\right)^3}-1, \quad
r_*= \left(\frac{8r_{g}r_c^2}{9}\right)^{1/3},
\label{eq:vain}
\end{equation}
and the Schwarzschild radius
\begin{equation}
r_{g} = 2GM = \frac{8\pi G}{3} \delta \rho_0 r_s^3 .
\end{equation}
%
The radius $r_*$ is called the Vainshtein radius. For
$r_s < R \ll r_*$, we obtain $g(R) \propto R^{-3/2}$, so
$\phi(R) \propto R^{1/2}$ +const. This means that for test bodies the
correction to Newtonian forces around the source vanishes in the limit
$r_s \ll r_*$, and thus Einstein gravity is recovered
\cite{Lue:2004rj}. This condition is satisfied if
$r_{g} \gg 9 r_s^3 /(8r_c^2)$ which is the case for stars and even the
  Earth and the Moon as long as we set $r_c$ to be a cosmological scale.
Equivalently, for spherically symmetric systems the Vainshtein mechanism
is active whenever the mean enclosed overdensity $3M/4\pi R^3$ is
greater than
%
\begin{equation}
  \rho_{\rm th} = \frac9{8(H_0 r_c)^2} \rho_{\rm cr},
\end{equation}
%
where $\rho_{\rm cr}$ is the critical density.

We can in fact obtain the full solution for $\phi$ in closed form
\cite{Schmidt:2009yj}. For $R \ge r_s$, defining $x\equiv R/r_*$ and
$A=3  r_*^2/(4r_c^2)$, we can express the solution of
Eq.~(\ref{eq:phisph2}) in terms of the hypergeometric function,
%
\begin{equation}
 \phi_{\rm ex}(x) = C_1 + \frac{A}{2}f(x),\qquad {\rm for}\,\, x \geq
  x_s,
 \label{eq:exsol}
\end{equation}
%
where $C_1$ is an integral constant, $x_s\equiv r_s/r_*$,  and
%
\begin{equation}
f(x) \equiv x^2\left[
  {}_2F_1\left(-\frac{1}{2},-\frac{2}{3};\frac{1}{3};-\frac{1}{x^3}\right)-1\right].
\end{equation}
%

On the other hand, in the internal region, $R<r_s$, the solution of
Eq.~(\ref{eq:phisph2}) is
%
\begin{equation}
 \phi_{\rm in}(x) = \frac{A}{2}x^2g(x_s) +C_2, \qquad {\rm for}\,\, x<x_s. \label{eq:insol}
\end{equation}
%
Note that in this case, the two pieces of the nonlinear terms combine and imply
%
\begin{equation}
\left[ (\nabla^2 \phi_{\rm in})^2 -
    (\nabla_i \nabla_j \phi_{\rm in})^2 \right]
    = {2 \over 3} (\nabla^2 \phi_{\rm in})^2.
\end{equation}
%
More generally, for any $\phi_A$
%
\begin{equation}
N[\phi_A,\phi_{\rm in}] = \frac{2}{3} r_c^2 \nabla^2 \phi_A \nabla^2\phi_{\rm in} .
\label{eq:Napprox}
\end{equation}
%
This relation is specific to the interior of a top hat $\phi_{\rm in}$
but nonetheless will be useful when approximating the nonlinear term.

The inner and external solutions should be continuous at $x=x_s$.
Without loss of generality, we can take $C_1=0$ due to the shift
symmetry of the scalar field equation of motion. Then we obtain the
solution for a single source, 
$\phi(x)=\phi_{\rm in}(x)\Theta(x_s-x)+\phi_{\rm ex}(x)\Theta(x-x_s)$ with 
%
\begin{equation}
 \phi_{\rm ex}(x) = \frac{A}{2}f(x), \quad
 \phi_{\rm in}(x) = \frac{A}{2}\left[f(x_s)+ g(x_s)(x^2-x_s^2)\right].
\label{eq:exin2}
\end{equation}
%
For $R\ll r_*$ and $r_s \ll r_*$, these solutions can be approximated as
%
\begin{eqnarray}
 \phi_{\rm ex}(R) &\approx &{\sqrt{2} r_{g}^{1/2} \over r_c}   R^{1/2} +A \frac{\Gamma[-2/3]\Gamma[7/6]}{\sqrt{\pi}}, \nonumber\\
  \phi_{\rm in}(R) &\approx &{r_{g}^{1/2} \over 2 \sqrt{2} r_c r_s^{3/2}} R^2 +
  {3r_{g}^{1/2} r_s^{1/2} \over 2 \sqrt{2} r_c }+A \frac{\Gamma[-2/3]\Gamma[7/6]}{\sqrt{\pi}} ,
\label{eq:exin2approx}
\end{eqnarray}
%
with a Laplacian of
%
\begin{equation}
\nabla^2 \phi_{\rm ex} (R) = {3 r_{g}^{1/2} \over 2 \sqrt{2} r_c} R^{-3/2}, \quad
\nabla^2 \phi_{\rm in} (R) = {3 r_{g}^{1/2} \over \sqrt{2} r_c r_s^{3/2}}.
\end{equation}
%
That the radial dependence of these solutions is proportional to
$r_g^{1/2} \propto M^{1/2}$ is an indication that mass sources do not
linearly superimpose within the Vainshtein radius. 
On the other hand in the opposite limit of $R \gg r_* \gg r_s$,
%
\begin{equation}
\phi(R) \approx  - \frac{r_g}{3R}\left( 1 - \frac{r_g r_c^2}{18R^3} \right).
\end{equation}
%
Leading order linearity in $r_g \propto M$ implies that mass sources do superimpose in this limit.
Since the leading order term in the Laplacian $\nabla^2\phi=0$, residual effects go as
%
\begin{equation}
\nabla^2\phi(R) \approx \frac{2}{9} \frac{r_g^2 r_c^2}{R^6}.
\label{eqn:Msquared}
\end{equation}
%
These approximate forms will be  useful in constructing scaling relations and boundary conditions for the two-body problem.

\section{Two-body problem}
\label{sec:superposition}

The brane bending field of a single body derived in the previous section
suffices to study the motion of test particles around that body.   For
test particles, there is a universal anomalous precession rate,
dependent only on the crossover scale $r_c$, that can be used to test
the Vainshtein mechanism \cite{Lue:2002sw,Dvali:2002vf}.
However for most realistic orbiting bodies, their own Vainshtein radius
is too large for them to be considered test particles.   This is, in
particular, true for the  Earth-Moon system where the orbit of the Moon
is well within its own Vainshtein radius. 

Through the nonlinearity of the Vainshtein effect, the orbiting body's
self-field interferes with that of the central body and can in principle
affect its orbit. Indeed the scaling of the single-body field as
$\sqrt{M}$ is an indication of the nonsuperimposability of solutions
within the Vainshtein radius. Two sources will add as
$\sqrt{M_A+M_B}$ for distances from the center of mass much larger than
the separation. 

In this section, we consider a two-body problem such as the Earth-Moon
system to study nonsuperimposability of solutions.  
We begin in Sec.~\ref{subsec:modelparams} with the parametrization of
the two-body 
system in terms of the physical scales in the problem.  We describe how
screening operates directly on second derivatives of the field and
indirectly on first derivatives, or   average forces in 
Sec.~\ref{sec:expectations}.   We examine the geometry of screening in 
Sec.~\ref{sec:NAB} and introduce our  screening statistics and their
scaling properties in Sec.~\ref{sec:Q2}.

\subsection{Model parameters}
\label{subsec:modelparams}

\begin{figure}[!ht]
  \centering{
  \includegraphics[width=4cm]{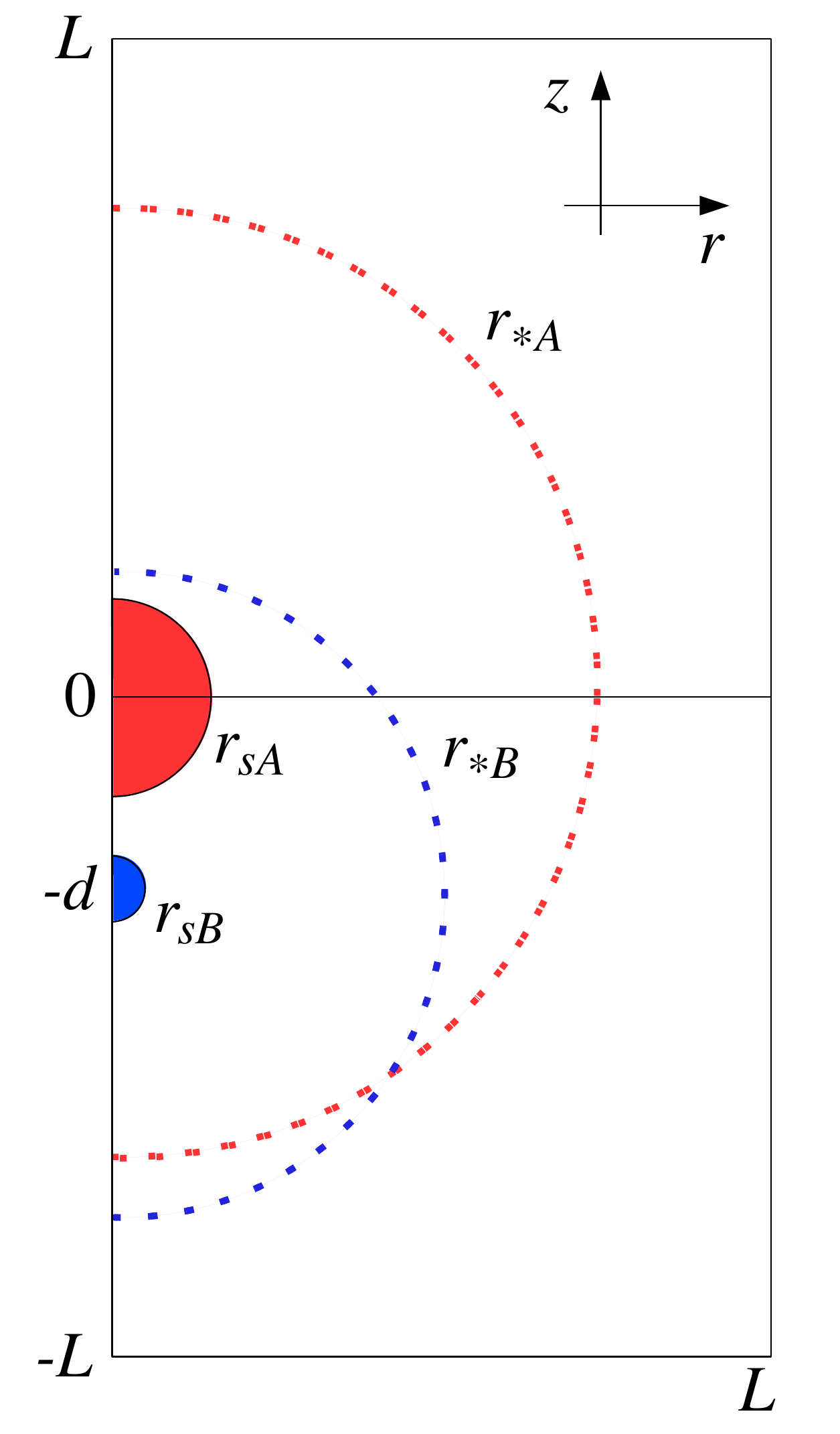}
  }
  \caption{Schematic illustration of the two body problem (not drawn to
scale). Physical radii are shown by the solid lines whereas the
 Vainshtein radii, $r_{*A}$ and $r_{*B}$ are shown in dashed lines.
 Boundary conditions are set at $r=L$ and $z=\pm L$ with vanishing
 deviations from superposition of  $A$ and $B$. 
  }
  \label{fig:compdom}
\end{figure}

Given spherically
symmetric masses, the system has axial symmetry and  so
we use cylindrical coordinates $(r, \theta, z)$. We assume that the 
two-bodies, denoted as $A$ and $B$, are separated from each other by $d$,
and they are located at $(r,z)=(0,0)$ for body $A$ and $(r,z)=(0,-d)$
for body $B$. We denote their respective Schwarzschild and physical
radii as $r_{gA}, r_{gB}$ and  $r_{sA}, r_{sB}$. In the following we
assume body $A$ is heavier than body $B$. 
The schematic illustration of the two-body set up is shown
in Fig.~\ref{fig:compdom}.

In our numerical solutions below, we choose parameters that reflect the
Earth (``E") and Moon (``M") where possible. In the actual Earth-Moon
system $d=3.8\times 10^{5}$ km and choosing $r_c \sim c H_0^{-1}$ as
appropriate for cosmologically motivated models would require a $10^8$
dynamic range between the Vainshtein radius of the Earth and the
separation $d$ since 
%
\begin{equation}
 \frac{cH_0^{-1}}{d} = 3.3\times 10^{17}\left(\frac{0.73}{h}\right).
\end{equation}
%
Furthermore between the separation scale $d$ and the body radii are
several more orders of magnitude
%
\begin{eqnarray}
r_{sE} = 6378\,{\rm km} = 1.7\times 10^{-2}d,\nonumber\\
r_{sM} = 1737\,{\rm km}=4.6\times 10^{-3}d.
\label{eq:actual_rs}
\end{eqnarray}
%
On the other hand, we shall see that the most important properties of
the two-body solution can be expressed as a function of the mass ratio
$M_B/M_A$, whereas other properties can be inferred by examining the
scalings of results with $r_c/d$ and $r_s/d$.

Our fiducial choice will therefore be to take $d$ and masses, or equivalently
Schwarzschild radii, from the  Earth-Moon system
%
\begin{equation}
\begin{aligned}
 \frac{r_{gA}}{d}&= 2 \frac{GM_{\rm E}}{d}= 2.3\times 10^{-11}, \\
 \frac{r_{gB}}{d}&\equiv  2 \frac{GM_{\rm M}}{d} =
 \frac{r_{gA}}{d} \left(\frac{M_M}{M_{E}}\right).
\end{aligned}
\end{equation}
%
Since we are interested in effects around body $B$ we fix $r_{sA}/d=0.3$
and explore the dependence on $r_{sB}$.  Likewise we explore the dependence of results
on $r_{c}/d$. 
 Our fiducial choices for these parameters  are listed in Table.~\ref{tab:fidparam}.
\begin{table}[!ht]
  \begin{tabular}{|c|c|}
\hline
  $M_B/M_A$ & 1/80\\
 $r_{gA}/d$  & $2.3 \times 10^{-11}$ \\
   $r_c/d$     & $10^8$  \\ 
 $r_{sA}/d$  & 0.3 \\
 $r_{sB}/d$  & 0.1 \\
 \hline
   $r_{gB}/d$  & $2.8 \times 10^{-13}$ \\
 $r_{*A}/d$  &  $58.91$ \\ 
 $r_{*B}/d$  & $13.67$ \\ 
 \hline
 $L/d$ & 286.7\\
 $h/d$ & 0.0125 \\
\hline
  \end{tabular}
  \caption{The fiducial parameters for the two-body problem where  $M_B/M_A$ and
  $r_{gA}/d$ mimic the Earth-Moon system.
  The top set represents the primary parameters whose individual
 variation we explore below.  The middle set represents derived
 parameters.  The last two parameters are numerical and give the box and
 grid size (see the Appendix).}
  \label{tab:fidparam}
\end{table}

It is useful both numerically and analytically to express the two-body problem
in terms of deviations from the superposition principle.
Denoting the
full solution as
%
\begin{equation}
\phi =\phi_A + \phi_B + \phi_\Delta,
\end{equation}
%
where $\phi_A$ and $\phi_B$ are the solutions to the single-body equations,
the field equation for the deviation $\phi_\Delta$ is
%
\begin{equation}
3 \nabla^2 \phi_\Delta + N[\phi_\Delta,\phi_\Delta]
+ 2 N[\phi_A+\phi_B,\phi_\Delta] = - 2N [\phi_A,\phi_B].
\label{eqn:phideltaeom}
\end{equation}
%
The interference term $N [\phi_A,\phi_B]$
can be considered as the source of deviations from superposition.

We solve Eq.~(\ref{eqn:phideltaeom}) numerically with boundary conditions
\begin{equation}
\phi_{\Delta}(L,z)= \phi_{\Delta}(r,-L)=\phi_{\Delta}(r,L)=0,
\end{equation}
corresponding to the case $L \gg r_{*A}$ where we can superpose
single-body solutions.  
Finally the equations are solved by finite difference on an
inhomogeneous grid with minimum spacing $h/d$ (see the Appendix). 

\subsection{Screening}
\label{sec:expectations}

Before turning to numerical solutions of the two-body system it is useful to
examine the structure of Eq.~(\ref{eqn:phideltaeom}) and develop an analytic
intuition for the results.
In particular, Eq.~(\ref{eqn:phideltaeom}) admits screening
solutions where the field of body $B$ suppresses some aspect of the
field of body $A$ around itself and vice versa.

To see this screening behavior note that both
$\phi_\Delta = -\phi_A$ and $\phi_\Delta=-\phi_B$ are solutions to this
equation {\it outside} of the sources where
$3\nabla^2 \phi_{A,B}+ N[\phi_{A,B},\phi_{A,B}]=0$.
In particular, around body $B$ we might expect $\phi_\Delta = -\phi_A$
such that it cancels the effect of body $A$. However these source-free
solutions would not match the boundary conditions at the sources
themselves and so what occurs in reality is more complicated.

Screening really occurs in the second derivatives of the field.
Note that we are also free to add a pure gradient to any of these
source-free solutions, e.g.
%
\begin{equation}
\phi_{\Delta} = -\phi_A + {\bf C}\cdot {\bf x} + D
\end{equation}
%
and still solve the equations.  Thus we might expect that screening
operates by replacing $\phi_A$ with a version of itself that is
linearized over some region of influence around body $B$ that can be much larger
than the physical size of the body (cf.~\cite{Hui:2009kc}).

We start with the simple expectations from approximating the nonlinear
term with the Laplacian in Eq.~(\ref{eq:Napprox}). In this approximation
the more general screening expectation becomes
%
\begin{eqnarray}
 \nabla^2 \phi_{\Delta}
 \approx
 \begin{cases}
   -\nabla^2 \phi_A, & \nabla^2\phi_B \gg \nabla^2 \phi_A\\
 -\nabla^2 \phi_B, &  \nabla^2\phi_B \ll \nabla^2 \phi_A
 \end{cases},
\end{eqnarray}
%
with the constants ${\bf C}$ and $D$ providing the appropriate matching of
the regimes.  Note that if $\phi_A$ is already nearly linear around body $B$,
as is the case for a distant source \cite{Hui:2009kc} we expect no
self-field effect on the motion of body $B$.
On the other hand we know that there must be a near field effect on
forces between the bodies: without screening the force from body $A$ on
body $B$ 
%
\begin{equation}
F_{AB} \propto M_B \sqrt{M_A},
\end{equation}
%
whereas that from body $B$ on body $A$
%
\begin{equation}
F_{BA} \propto -M_A \sqrt{M_B}.
\end{equation}
%
This violation of Newton's third law would cause momentum nonconservation
in the joint system.

To get a rough sense for the size and scaling of near field effects we
can replace $\phi_A \propto \sqrt{R}$ with a linearization of itself
across the region on the $z$ axis where $\nabla^2 \phi_B > \nabla^2
\phi_A$ 
%
\begin{equation}
\left|{z_{\rm min} \over d}\right|  = {1 \over 1+(M_B/M_A)^{1/3}},
\quad
\left|{z_{\rm max} \over d}\right|  = {1 \over 1-(M_B/M_A)^{1/3}},
\end{equation}
%
or
%
\begin{equation}
\phi_{\Delta}(0,z) = -\phi_A(0,z) + \frac{ \phi_A(0,z_{\rm max}) -
 \phi_A(0,z_{\rm min}) }{z_{\rm max}-z_{\rm min}} (z-z_{\rm min})+
 \phi_A(0,z_{\rm min}) . 
\end{equation}
%
In this crude approximation, the gradient $\partial_z \phi_A (0,-d)$ is reduced by
%
\begin{equation}
\frac{ \partial_z \phi_\Delta}{\partial_z \phi_A} (0,-d) \sim -\frac{3}{8} \left( \frac{M_B}{M_A} \right)^{2/3}
+ {\cal O}\left( \frac{M_B}{M_A} \right)^{4/3}.
\label{eqn:crude}
\end{equation}
%
independently of the physical size of the bodies and the value of $r_c/d \gg 1$.
For the Earth-Moon mass ratio this is a $\sim 2\%$ correction of the gradient even though the
second derivative is screened across a much larger range, 
$\Delta z/d \sim 0.5$,  than the physical size of the Moon 
$\Delta z/d \sim 0.005$. Nonetheless the dependence on the mass ratio
represents an apparent violation of the equivalence principle. 

Note that momentum conservation in the joint system would imply that at
body $A$ there is near complete screening of the force from body $B$ 
%
\begin{equation}
\frac{ \partial_z \phi_\Delta}{\partial_z \phi_B}(0,0) \approx -1 + \sqrt{M_B \over M_A}
\left[ 1 + \frac{ \partial_z \phi_\Delta}{\partial_z \phi_A} (0,-d)\right]\label{eq:momcons}
\end{equation}
%
for $r_{sA},r_{sB}\ll d$. As a check of our numerical results we will
examine
%
\begin{equation} 
\frac{F_{AB}}{F_{BA}}= \frac{M_B}{M_A} \frac{ \left(\partial_z \phi_A +
			\partial_z\phi_\Delta\right) |_{0,-d}}
{\left(\partial_z \phi_B + \partial_z\phi_\Delta\right) |_{0,0}} 
\label{eqn:3rd}
\end{equation}
%
to determine how well Newton's third law $F_{BA}/F_{AB}=-1$ is satisfied.

Note that in the opposite limit $d \ll r \ll r_{A*}$,  where the sources add as
$\sqrt{M_A+M_B}$, we know that the field of body $A$ screens that of
body $B$ more directly.   The $\phi_\Delta$ field here
is simply the difference between the joint and individual sources
%
\begin{equation}
 \phi_\Delta \propto  \sqrt{M_A+M_B} -\sqrt{M_A}-\sqrt{M_B} \approx -\sqrt{M_B},
 \label{eqn:sqrtM}
\end{equation}
%
for $M_B \ll M_A$ implying
%
\begin{equation}
 \phi_\Delta \approx -\phi_B.
\end{equation}
%

While these considerations provide a qualitative guide to results,
the specific form of the second derivatives in $N[\phi_A,\phi_B]$ lead
to important consequences for the geometry of the screening around body
$B$ which we shall now consider.

\begin{figure}
\includegraphics[width=0.5\textwidth]{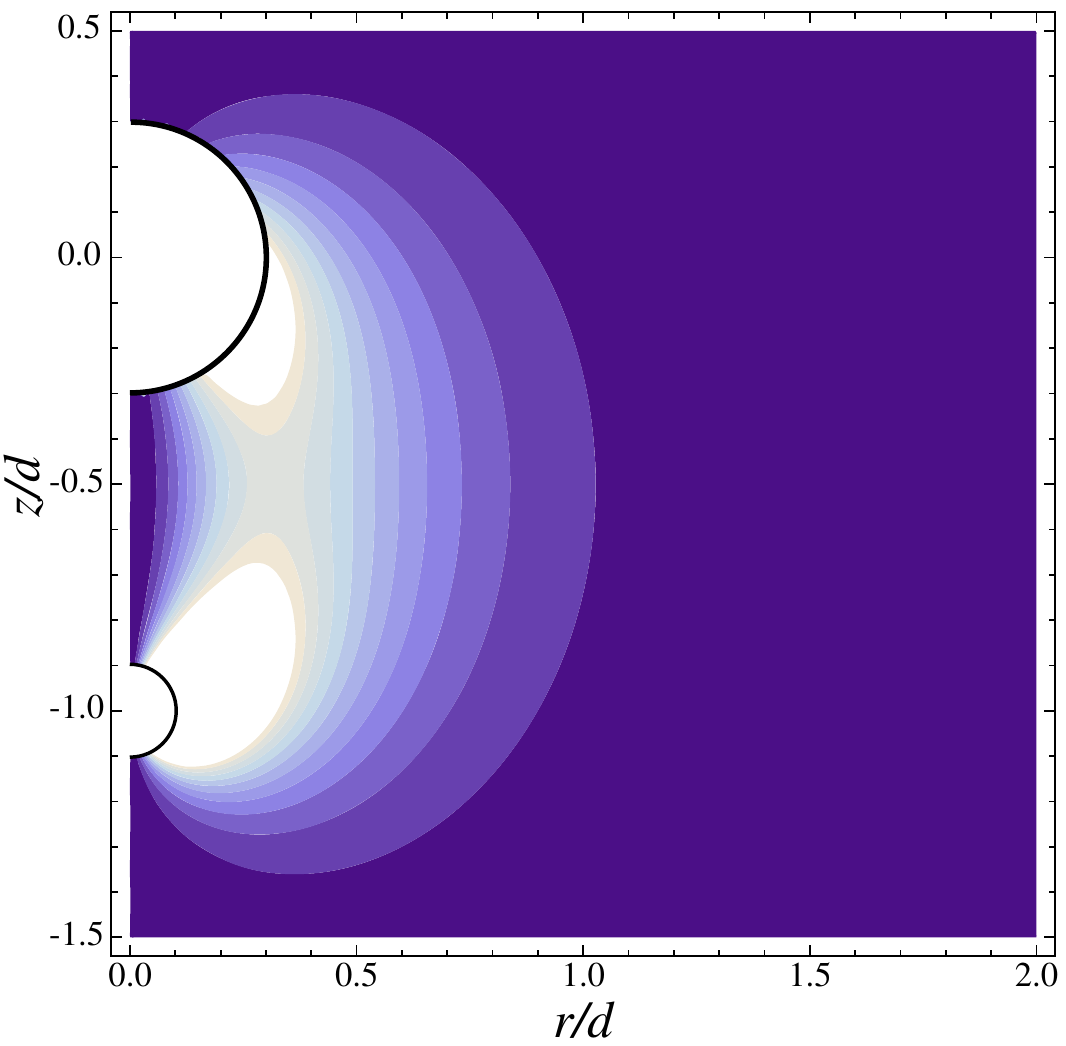}
\caption{Two-body nonlinearity source function $ F[r/d,z/d] \propto
 N[\phi_A,\phi_B]$ with $A$ at $(0,0)$ and 
$B$ at $(0,-1)$ with parameters of the fiducial model (see
 Table~\ref{tab:fidparam}).  The region interior to the bodies, within
 the semicircles, is not shown.} 
\label{fig:sources}
\end{figure}

\subsection{Toroidal geometry}
\label{sec:NAB}

In order to understand the geometry of screening around body $B$, let us
examine the source to the $\phi_\Delta$ field in
Eq.~(\ref{eqn:phideltaeom})
%
\begin{equation}
2 N[\phi_A,\phi_B]  = {9 \over 2} r_c^2  {\sqrt{G M_A} \sqrt{G M_B } \over d^3}  F[r/d,z/d],
\end{equation}
%
where the geometry is determined by a universal function in cylindrical coordinates scaled to the separation $d$
%
\begin{equation}
F[\tilde r, \tilde z] =
 \frac{\tilde r^2}{(\tilde r^2 + \tilde z^2)^{7/4}(\tilde r^2 + (\tilde z + 1)^2)^{7/4}}.
\label{eqn:exteriorsource}
\end{equation}
%
This function is
plotted in Fig.~\ref{fig:sources}.   Instead of the roughly spherical geometry that the arguments based on the Laplacian would predict, the true interference of the self-field of $B$ on that of $A$ is toroidal around body $B$.    The basic reason for this geometry
is that along the $z$ axis the cross terms cancel given the difference structure in
Eq.~(\ref{eq:defN}).

 On the other
hand, interior to body $B$, the nonlinear source becomes
\begin{equation}
 2 N[\phi_A,\phi_B]  = {6r_c^2  \sqrt{G M_A} \sqrt{G M_B } \over r_{sB}^{3/2} (r^2+z^2)^{3/4}}
\end{equation}
and does not vanish for $r=0$ but rather approaches a constant for $d \gg r_{sB}$.
Note that for a small physical size $r_{sB}$, $r^2+z^2 \approx d^2$ in
the interior.  The interference term is nearly constant and approximates an effective
density of $\sqrt{\rho_B \rho_{A,{\rm eff}}}$  where
$\rho_{A,{\rm eff}}=M_A/(4\pi d^3/3)$. Unlike the true density,
$2N[\phi_A,\phi_B]$ does not vanish in the exterior but has a jump in
value at the $r_{sB}$ boundary.  We shall see that this jump causes a similar discontinuity in the
second derivatives of $\phi_\Delta$ at the boundary.

Finally, although $2N[\phi_A,\phi_B]$ peaks at the
bodies themselves (saturated in Fig.~\ref{fig:sources}), this does not
necessarily mean that the relative impact on the joint field peaks
there.  The single-body Laplacians also peak there and we must examine
deviations with respect to those fields in the results that follow. 

\subsection{Screening statistics}
\label{sec:Q2}

It is therefore useful to introduce our primary, or Laplacian, screening
statistic for the relative impact of the field of body $B$ on that of
$A$ 
\begin{equation}
Q_2(r,z) = \frac{\nabla^2\phi_\Delta}{\nabla^2\phi_A}.\label{eq:defQ2}
\end{equation}
If screening is complete, $Q_2 \rightarrow -1$.

It is again useful to derive rough analytic scalings for the screening statistic in various limits.
 When either $A$ or $B$
dominate in the second derivatives we expect that the second derivatives
of the correction field $\phi_\Delta$ is small compared with the
dominant one and screens the subdominant one.   We therefore expect the
field equation (\ref{eqn:phideltaeom}) to be satisfied approximately by
dropping terms nonlinear in $\phi_\Delta$.   Furthermore, the
$3\nabla^2\phi_\Delta$ terms is small compared with the nonlinear terms
deep within the Vainshtein radius and so
\begin{equation}
  N[\phi_A+\phi_B,\phi_\Delta] \approx - N [\phi_A,\phi_B].
\label{eqn:phideltaapp}
\end{equation}
\begin{figure}[t]
  \centering{
  \includegraphics[width=6in]{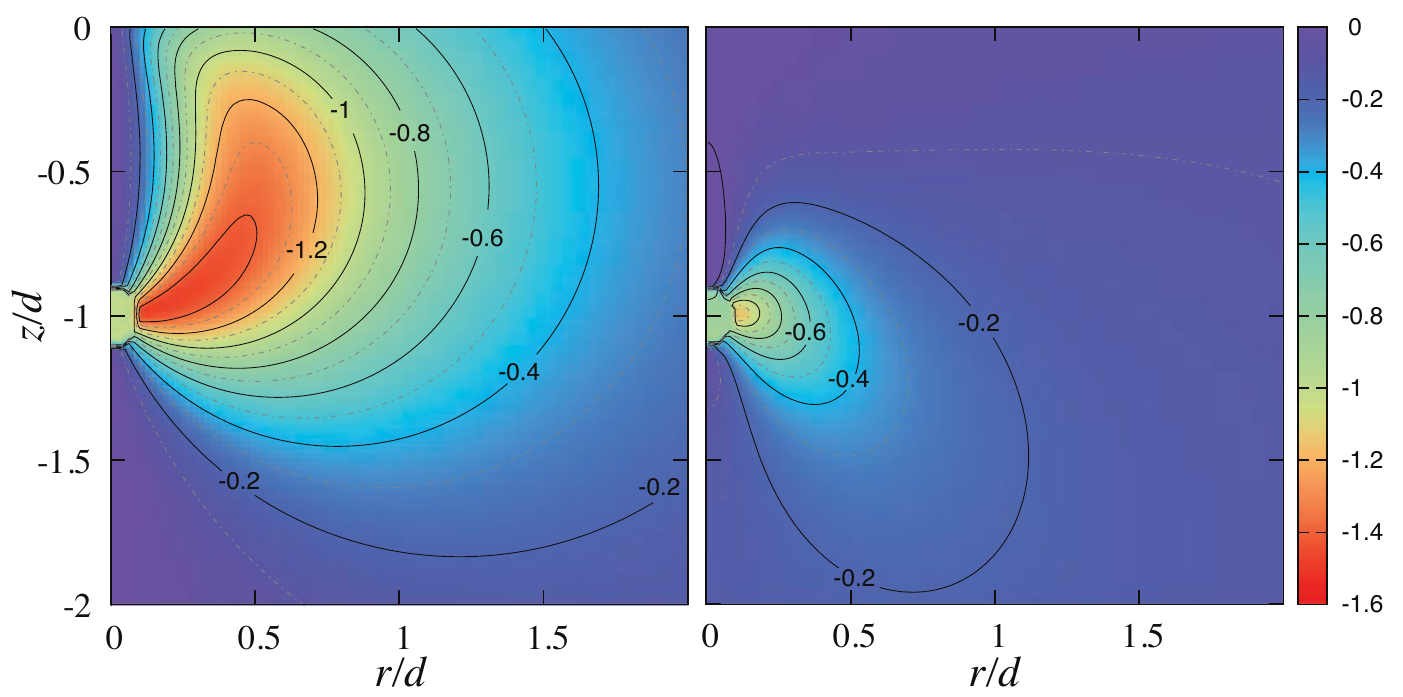}
  }
  \caption{Laplacian-screening statistic $Q_2=\nabla^2\phi_\Delta /\nabla^2\phi_A$ (left: analytic; right: numerical).
 The red areas indicate a large relative deviation from the superposition
 solution in the Laplacian of the field.  In the interior of body $B$ screening of the Laplacian
  of body $A$ is nearly complete whereas in the exterior there is a toroidal region of
 mutual screening.   The analytic description is in good qualitative agreement near
 body $B$.
  Parameters are the fiducial choice of Table~\ref{tab:fidparam}.  }
  \label{fig:P1P2_cont}
\end{figure}
We can further approximate the left-hand side of Eq.~(\ref{eqn:phideltaapp}) using Eq.~(\ref{eq:Napprox})
\begin{equation}
N[\phi_A+\phi_B,\phi_\Delta] \approx \frac{2}{3} r_c^2
     \nabla^2 (\phi_A + \phi_B)
     \nabla^2 \phi_{\Delta} .
    \label{eq:Napprox2}
\end{equation}
This relation is exact only for the interior field and for $\nabla^2\phi_B \gg \nabla^2\phi_A$ but gives a reasonable heuristic description elsewhere.
This approximation is useful in that it allows us to solve directly for $\nabla^2\phi_\Delta$, and hence
%
\begin{align}
Q_{2 \rm in}(r,z)
  & \approx - \left[ 1+
        \frac{1}{2} \left( \frac{M_A}{M_B} \right)^{1/2}
        \left(\frac{r_{ sB}^{2}}{r^2 + z^2}\right)^{3/4}
     \right]^{-1},
\label{eq:phiDI} \\
Q_{2 \rm ex}(r,z)
  & \approx -         \left\{ \frac{3}{2} \frac{d^2 r^2}{(r^2+z^2)[r^2+(z+d)^2]} \right\}
        \left\{ 1+
           \left(\frac{M_A}{M_B} \right)^{1/2}
           \left[\frac{r^2+(z+d)^2}{r^2 + z^2}\right]^{3/4}
        \right\}^{-1}.
\label{eq:phiDE}
\end{align}
%
The left-hand panel in Fig.~\ref{fig:P1P2_cont}  shows the results of $Q_2(r,z)$ obtained
from Eqs.~(\ref{eq:phiDI}) and (\ref{eq:phiDE}). 

In fact for small radius $r \ll d$, we can approximate $r^2+z^2 \sim d^2$. Then
on the $z=-d$ axis, $Q_2$ is approximately given by
\begin{align}
Q_{2 \rm in}(r,-d) &\approx  -\left[ 1 + \frac{1}{2} \left(\frac{M_A}{M_B} \right)^{1/2} \left(\frac{r_{sB}}{d}\right)^{3/2} \right]^{-1},
\label{eq:Q2in}
\\
Q_{2 \rm ex}(r,-d) &\approx -\frac{3}{2} \left[ 1 + \left(\frac{M_A}{M_B} \right)^{1/2} \left(\frac{r}{d}\right)^{3/2} \right]^{-1},
\label{Q2ex}
\end{align}
Note that saturation to $Q_2=-1$ in the interior increases with increasing $M_B/M_A$
and decreasing $r_{sB}$ as expected from the fact that the interior value of $\nabla^2\phi_B$
scales with these parameters.    We can infer from this scaling that for realistic
situations where $r_{sB} \ll d$ (see Eq.~\ref{eq:actual_rs}), $Q_2\approx -1$ for any $M_B<M_A$.  At the body surface, there is a jump to a maximum value
of $-3/2$.   We shall see from numerical results that this maximum value is only approximate since  Eq.~(\ref{eq:Napprox2}) is not strictly valid in this limit.

Nonetheless, the qualitative aspects of $Q_{2}$ indicate that the impact on forces is as a redistribution of force changes across a toroidal region around body $B$ rather than a linearization across a quasispherical one.  Since the volume of the regions are comparable, we expect the scaling behavior of Eq.~(\ref{eqn:crude}) to be roughly satisfied.
 To quantify this expectation, we define the force-screening statistic
%
\begin{equation}
Q_1(r)  =  \frac{\partial_z \phi_\Delta}{\partial_z \phi_A} \Big|_{z=-d} .
\label{eq:defQ1}
\end{equation}
%
Note that along $z=-d$ the gradient of the $\phi_B$ field is along the
$r$ direction and hence  its screening does not contribute to $Q_1$.
Again if screening of forces from $\phi_A$ is complete then
$Q_1\rightarrow -1$. 

For distances $R=\sqrt{r^2+z^2}$ from the bodies that are large compared
with the separation $d$, Eqs.~(\ref{eqn:sqrtM}) and (\ref{eqn:Msquared})
imply 
%
\begin{equation}
Q_2 \approx
\begin{cases}
-\sqrt{ \frac{M_B}{M_A}}, & d \ll R \ll r_{*A} \\
2\frac{M_B}{M_A}, & R\gg r_{*A}
\end{cases},
\label{eqn:farfield}
\end{equation}
%
which will also be useful in checking our numerical results.
In particular, $Q_2$ does not depend explicitly on $r_c$ aside from setting the transition scale $r_{*A}$.   In the
$R \gg r_{*A}$ regime this independence is due to the vanishing of $\nabla^2\phi_A$ to leading order.   Other statistics do not share this independence.
To see the more general dependence on $r_c$ note that the field equation for deviations from
superposition Eq.~(\ref{eqn:phideltaeom}) only has no explicit $r_c$
dependence when the nonlinear terms dominate
%
\begin{equation}
 N[\phi_\Delta,\phi_\Delta]
+ 2 N[\phi_A+\phi_B,\phi_\Delta] \approx - 2N [\phi_A,\phi_B]
\label{eqn:phideltaeom2}
\end{equation}
%
and hence fractional effects of $\phi_\Delta$ relative to $\phi_A$ or $\phi_B$ have no
$r_c$ dependence.
To see when this approximation is valid, take the opposite $r_c \rightarrow 0$ limit.
In that case,
%
\begin{equation}
3 \nabla^2 \phi_\Delta \approx - 2N [\phi_A,\phi_B] \propto r_c^2, \qquad r_c \rightarrow 0.
\label{eqn:phideltaeom1}
\end{equation}
%
The linear and nonlinear terms in Eq.~(\ref{eqn:phideltaeom}) become comparable  when
%
\begin{equation}
{\rm max}( r_c^2 \nabla_i\nabla_j \phi_B , r_c^2  \nabla_i\nabla_j \phi_A) \approx 1
\end{equation}
%
 which occurs for the typical distance $d$ around the bodies
 when $r_{*A} \approx d$.  Once $r_{*B} \gg d$, Eq.~(\ref{eqn:phideltaeom2})
 becomes valid and all statistics lose their  dependence on $r_c$ around the bodies.

\begin{figure}[b]
  \centering{
  \includegraphics[width=6in]{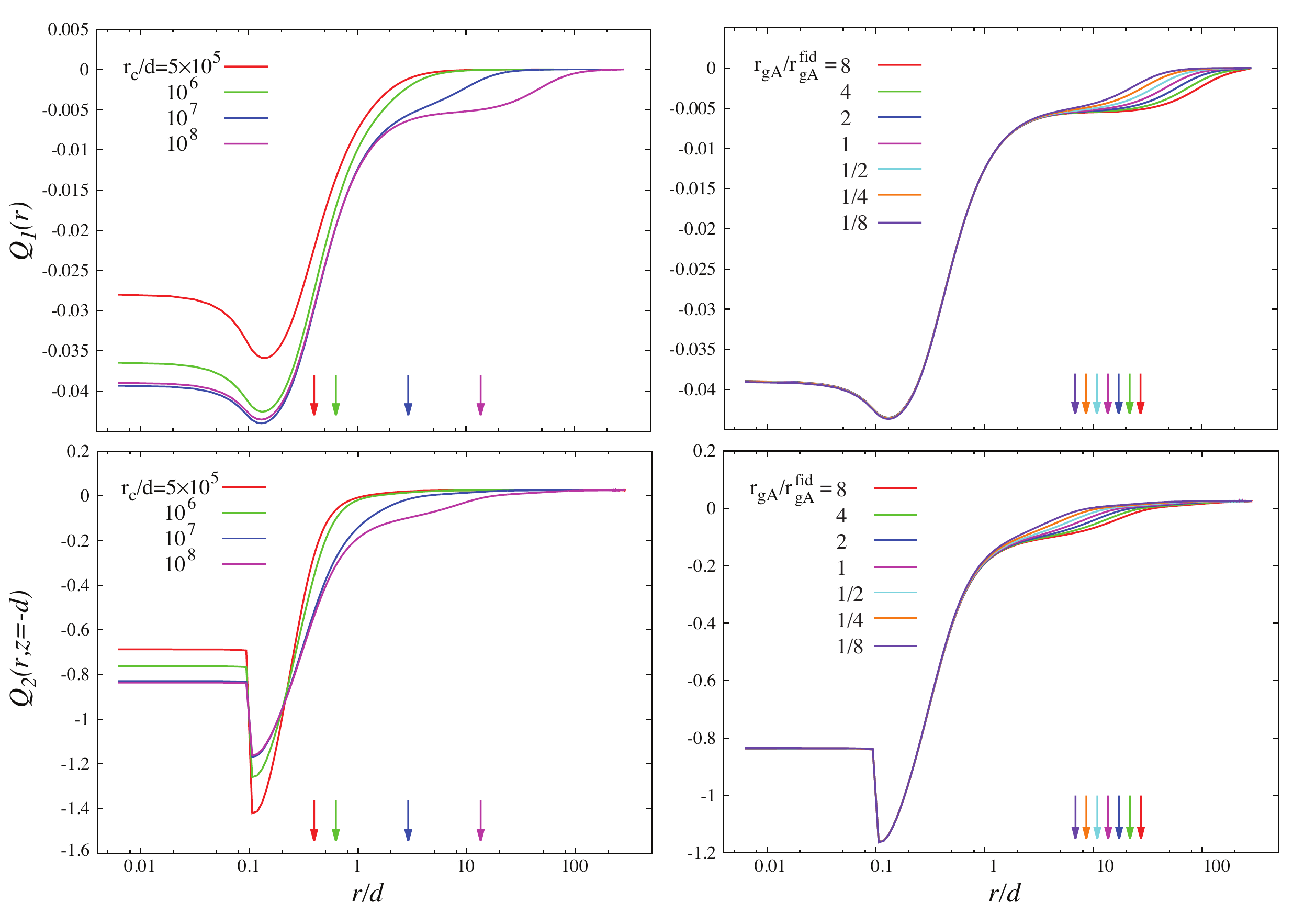}
  }
  \caption{Force ($Q_1$, top) and Laplacian ($Q_2$, bottom) screening
 statistics  along $z=-d$ as a function of the crossover scale $r_c$ (left)
 and absolute mass scale $r_{gA}=2GM_A$ (right).
 Other parameters including $M_B/M_A$
 are set to the fiducial choices of Table~\ref{tab:fidparam} here and in
 the following figures.  Arrows indicate the Vainshtein scale of body $B$, $r_{*B}$.
Near the body the screening statistics are independent of $r_c$ and mass scale if $r_{*B} \gg d$.}
  \label{fig:Q}
\end{figure}

\section{Results}
\label{sec:results}

In this section we present numerical solutions to the full nonlinear
equation and scaling relations based on them. The details of numerical
calculations are presented in the Appendix.

We begin by considering the fiducial parameter choices from
Table~\ref{tab:fidparam}. In Fig.~\ref{fig:P1P2_cont} (right)  we show a
2D contour plot of the Laplacian-screening statistic $Q_2$ of
Eq.~(\ref{eq:defQ2}). As expected from the analytic argument of the
previous section, $Q_2 \approx -1$ interior to  body $B$ indicating nearly full screening.
At the
surface of the body, it experiences a jump to $Q_2 < -1$ which then
extends to a toroidal region around the body.      In this toroidal region,
where the individual body Laplacians are comparable, $Q_2$ quantifies
both the screening effect of $B$ on $A$ and $A$ on $B$.  Comparison with Fig.~\ref{fig:P1P2_cont} (left)  shows that our crude analytic approximation of Eq.~(\ref{Q2ex}) captures many of the qualitative effects around body $B$.

To isolate the effect of screening of $B$ on $A$, it is useful to
examine the force-screening statistic $Q_1$ of Eq.~(\ref{eq:defQ1}).
The field of $B$ is purely radial around $B$ and $Q_1$ measures the
change in the gradient along the $z$ direction.  As shown in
Fig.~\ref{fig:Q} (top), $Q_1$ increases toward body $B$ and then
smoothly approaches a constant in the interior of the body.   For this
Earth-Moon-like system $Q_1(0) \approx -0.04$ or approximately double
the crude expectation from Eq.~(\ref{eqn:crude}).  The more dramatic
changes in $Q_2$ (see Fig.~\ref{fig:Q}, bottom) reflect changes in the
radial structure of the joint field.

The fiducial parameter choices reflect a crossover scale $r_c/d$  that is much smaller than the actual Earth-Moon system, a coupling strength $\beta$ set to unity by rescaling the masses $r_{g}/d$, and 
body sizes $r_s/d$ that are large compared with the separation. 
We therefore next test the dependence of our results
on these parameters before turning to the central dependence on the mass ratio $M_B/M_A$.   

In Fig.~\ref{fig:Q} (left), we show the dependence of the force and
Laplacian screening statistics on $r_c/d$. For $r_c \ll 10^{7} d$, the
screening statistics scales strongly with $r_c$ as expected from
Eq.~(\ref{eqn:phideltaeom1}).  This behavior saturates once 
$r_{*B} \gg d$.   In Fig.~\ref{fig:Q}, we show the $r_{*B}$ value
corresponding to $r_c$ with an arrow.   Indeed so long as 
$r_c \gg 10^7 d$ or equivalently $r_{*B} \gg d$ the results for both are
insensitive to $r_c$ near body $B$. As expected from
Eq.~(\ref{eqn:farfield}), for large $r_c/d$ there is an interval 
$d \ll r \ll r_{*A}$ where $Q_2 \approx -\sqrt{M_B/M_A}$. 
For $r\gtrsim r_{*A}$, $Q_2 \approx 2 M_B/M_A$ verifying that we have
taken a sufficiently large $L$ that the boundary condition
$\phi_\Delta=0$ is appropriate (see Sec.~\ref{appsec:tests} for an
explicit test).  

In Fig.~\ref{fig:Q} (right), we show the dependence on the absolute mass
scale or $r_{gA}/d$ at fixed $M_B/M_A$ and other fiducial parameters.
Recall that changing the mass scale can also be interpreted as changing
the parameter $\beta$ in the original field equation (\ref{eq:phievo}).
The only dependence of results on the mass scale is through its effect
on the Vainshtein scales and in the external far field limit relative to
the separation $d$. 

\begin{figure}[t]
  \centering{
  \includegraphics[width=6in]{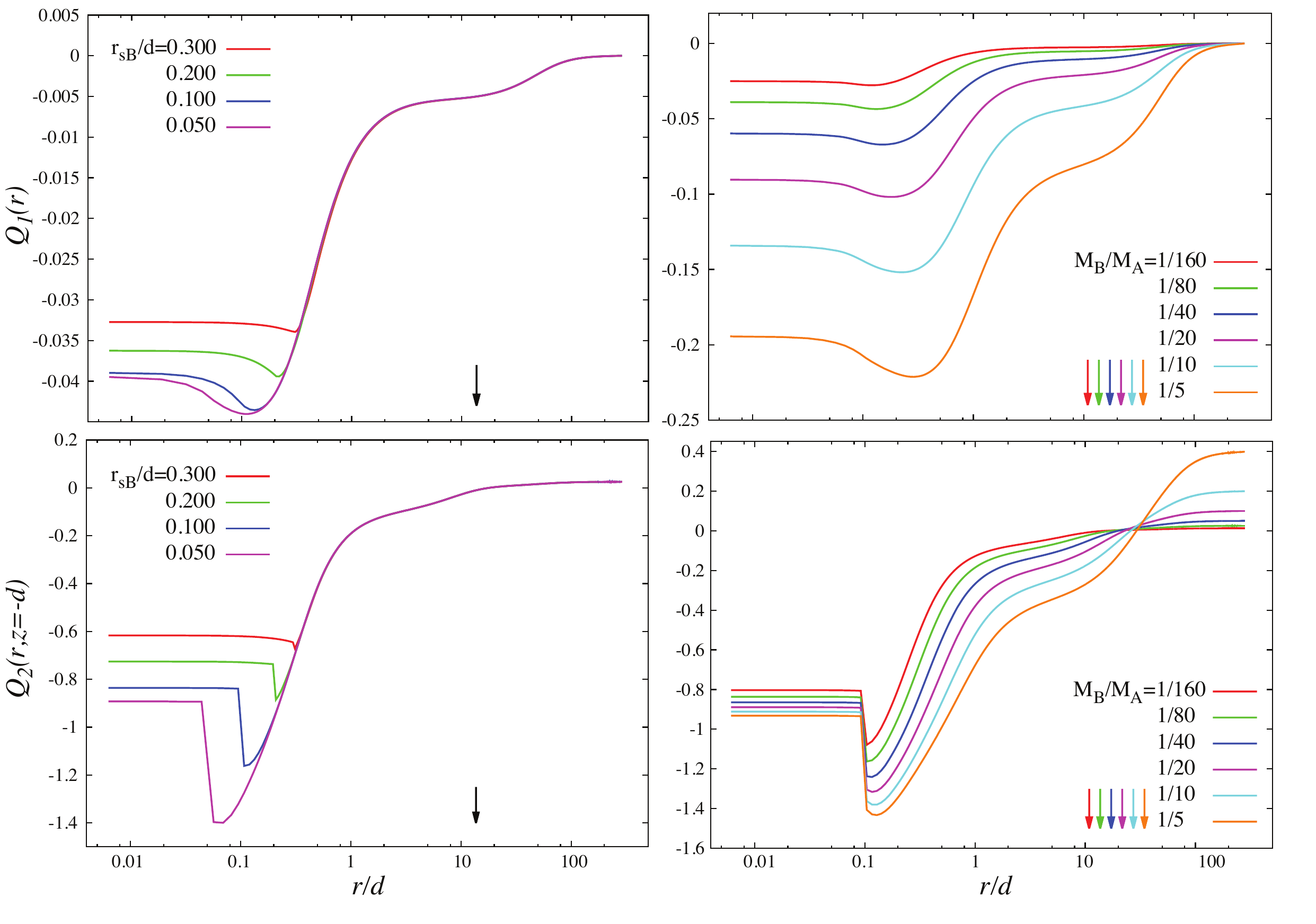}
  }
  \caption{Screening statistics as a function of  the size of body $B$ $r_{sB}$ (left) and mass ratio
 $M_B/M_A$ (right) as in Fig.~\ref{fig:Q}.  $Q_1$ and $Q_2$ are independent of $r_{sB}$ in the exterior $r>r_{sB}$ and converge to constant values for $r<r_{sB} \ll d$.
Likewise for $M_B/M_A$, $Q_2$ behaves according to analytic expectations for
$r\gg d$ and both $Q_1$ and $Q_2$ scale with the mass ratio in the interior $r<r_{sB}=0.1$.}  \label{fig:sizedep1}
\end{figure}

Next we consider the impact of the physical size of body $B$, $r_{sB}$ in Fig.~\ref{fig:sizedep1} (left).    As $r_{sB}$ shrinks, $Q_2$ interior to the body  approaches $-1$.  This is expected from our analytic expression, Eq.~(\ref{Q2ex}),
due to the fact that the maximum value
that $\nabla^2\phi_B$ attains is controlled by $r_{sB}$.   Once $\nabla^2\phi_B \gg \nabla^2\phi_A$ in the interior we expect results to become independent of $r_{sB}$.
For the fiducial parameters, $Q_2$ has not quite saturated whereas in $Q_1$ it has almost
reached its asymptotic value by $r_{sB}/d=0.1$.   In the exterior of body $B$,
 $Q_2$ drops increasingly below $-1$ as $r_{sB}/d$ decreases.    This large overshoot is not reflected in
$Q_1$ and hence represents the screening of the radial body $B$ field itself.   Again
since the maximum $\nabla^2\phi_B$ increases as $r_{sB}$ decreases, $\nabla^2\phi_\Delta$ increases relative to the constant $\nabla^2\phi_A$ as well for the screening of the $B$ field.  Since statistics in the exterior of body $B$ do not depend on its
size, there is likewise no dependence on the size of body $A$, $r_{sA}$ exterior to $A$.

We conclude that in the limit of $r_c/d\gg 1$ and $r_{sB}/d\ll 1$, the main
dependence of the screening variables on the system parameters is through
the mass ratio $M_B/M_A$.  This dependence is shown in
Fig.~\ref{fig:sizedep1} (right).    For  $Q_2$ interior to body $B$, 
raising the ratio increases the efficacy of screening body $A$ as
expected from Eq.~(\ref{eq:phiDI}).    In the exterior, it increases the
overshooting effect, also as expected.   For $Q_1$, screening in the
interior scales strongly with the mass ratio as expected from
Eq.~(\ref{eqn:crude}).   In both cases, in the far field limit $r \gg
d$, results scale according to the analytic expectations of
Eqs.~(\ref{eqn:sqrtM}), (\ref{eqn:Msquared}), and (\ref{eqn:farfield})
to excellent approximation.

\begin{figure}[!ht]
  \centering{
    \includegraphics[width=8cm]{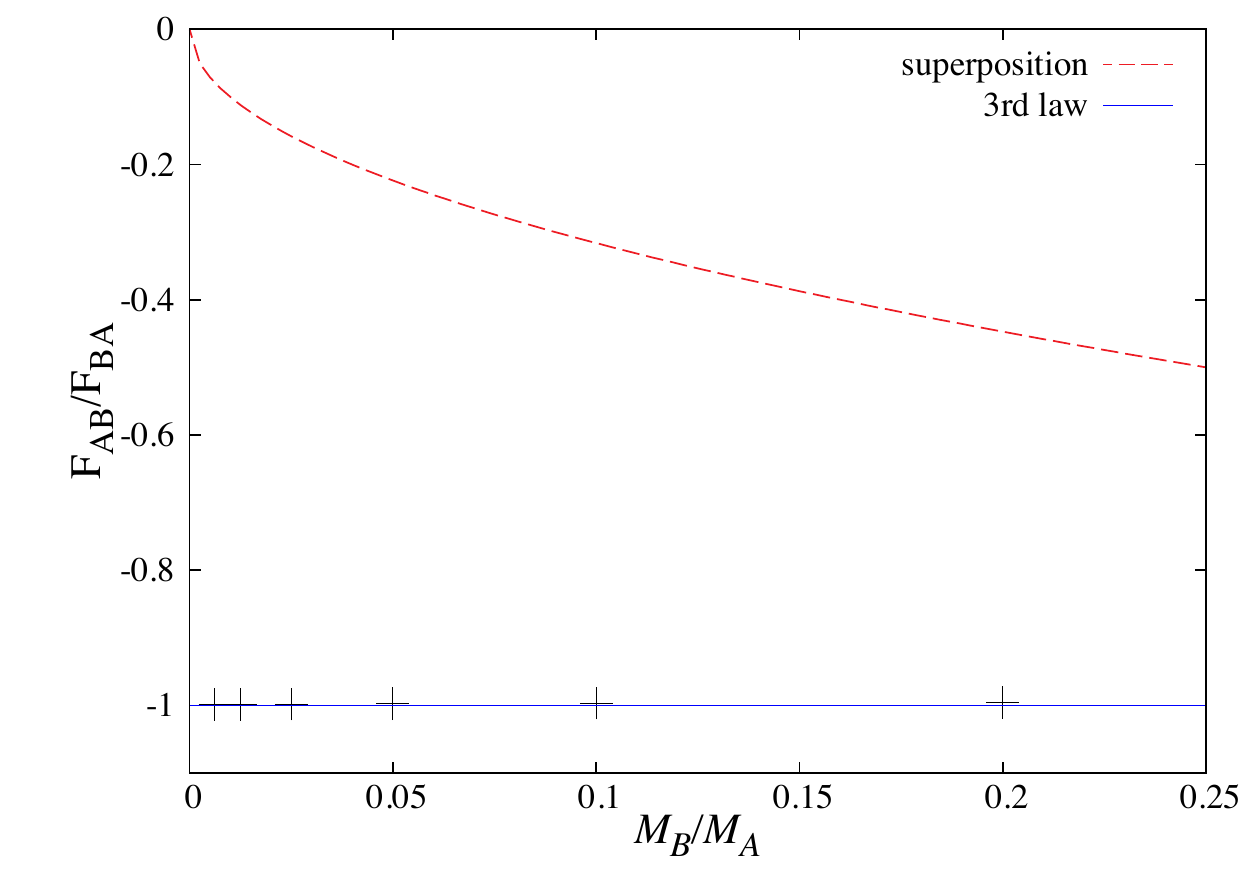}
  }
  \caption{Test of Newton's third law or momentum conservation.  If the fields of $A$ and $B$ superimpose, then the ratio of forces would be $-\sqrt{M_B/M_A}$ (red dashed line) and violate the third law (solid blue line).   Numerical results ($+$ points) show that the joint solution restores the third law through violation of the superposition principle $\phi_\Delta \ne 0$.
  }
  \label{fig:S29}
\end{figure}

As shown in Fig.~\ref{fig:S29}, these results for $Q_1$ near body $B$ are exactly what is required to restore Newton's third law and momentum conservation
[see Eq.~(\ref{eqn:3rd})].
Had the two fields superimposed, then the force ratio would be $-\sqrt{M_B/M_A}$ whereas our numerical results are consistent with $-1$.

\begin{figure}[!ht]
  \centering{
  \includegraphics[width=6in]{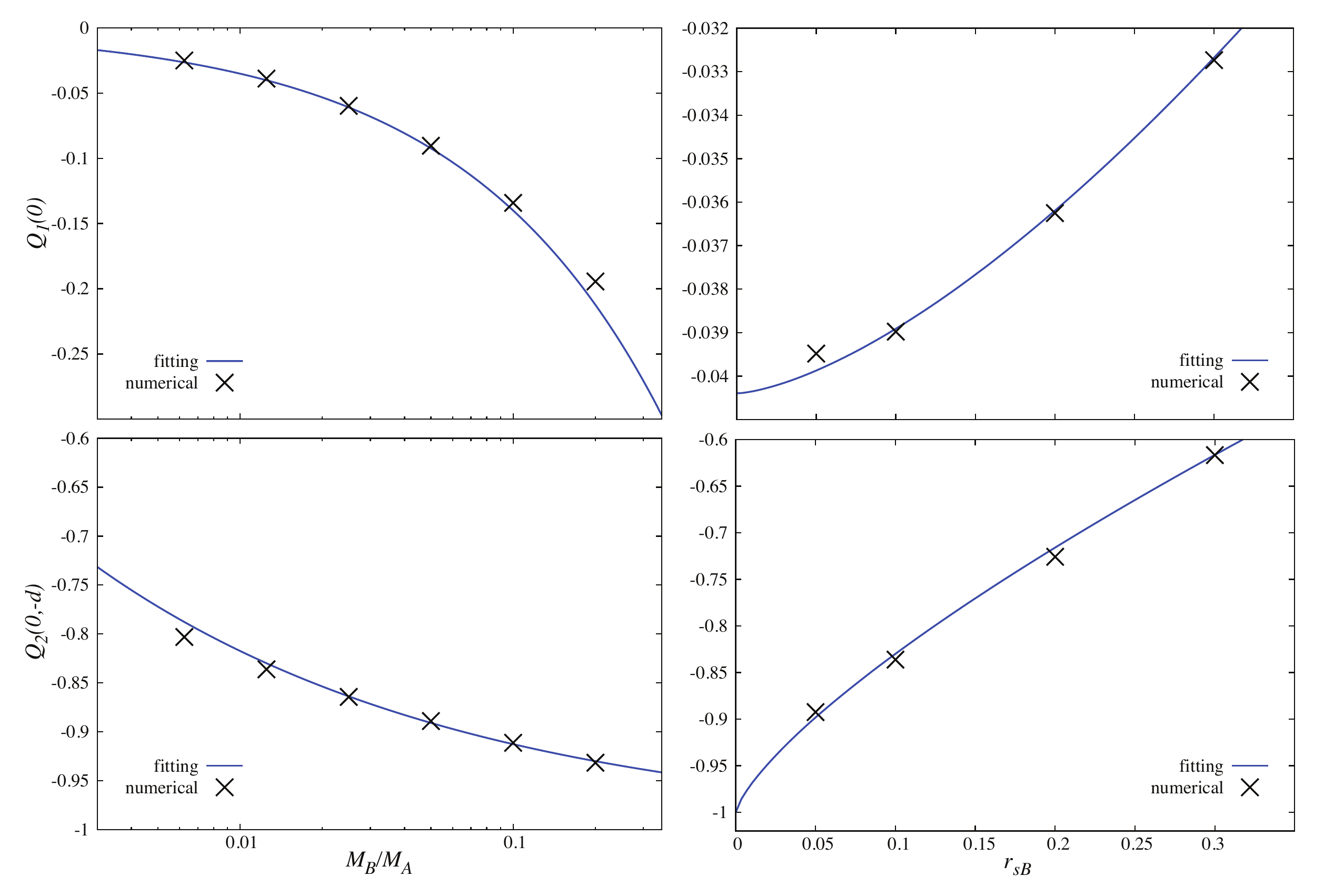}
  }
  \caption{Screening statistics at the center of body $B$, $Q_1(0)$ and $Q_2(0,-d)$ for various $r_{sB}$ and
 $M_B/M_A$ with other parameters held fixed to fiducial values.  We show numerical results as points ($\times$) as well as the
 fitting functions in Eqs.~(\ref{eqn:fitQ1}) and (\ref{eqn:fitQ2}).}
  \label{fig:S27}
\end{figure}

While our analytic approximations provide a qualitative description of our results, near body $B$ the geometry of the screening inhibits their accuracy.  
It is therefore useful to quantify $Q_1$ and $Q_2$ with empirical fits  at the center of body $B$.
 For $Q_1$, the numerical results can be described
by a scaling relation similar to Eq.~(\ref{eqn:crude}) but with a finite size correction (see Fig.~\ref{fig:S27})
%
\begin{equation}
Q_1(0) \approx -0.56 \left( \frac{M_B}{M_A} \right)^{0.6} \left[ 1 - 0.13 \left(\frac{M_A}{M_B}\right)^{1/2}
\left( \frac{r_{sB}}{d} \right)^{3/2} \right],
\label{eqn:fitQ1}
\end{equation}
%
for $M_B/M_A \lesssim 0.1$ and when the correction in brackets is small.
In particular, we can extrapolate this fit to the physically interesting limit
where $r_{sB}/d \ll 1$, $Q_1(0) \approx -0.56 (M_B/M_A)^{0.6}$.

For $Q_2$, our results are consistent with approaching $-1$ in the physically relevant limit
\begin{equation}
Q_2(0,-d) \approx -1 +0.23 \left( \frac{r_{sB}}{d} \right)^{0.74} \left( \frac{M_B}{M_A} \right)^{-0.32}.
\label{eqn:fitQ2}
\end{equation}
In practice, the approach is somewhat slower than that predicted by
Eq.~(\ref{eq:phiDI})  due primarily to inaccuracy in the replacement of second derivatives with Laplacians in Eq.~(\ref{eq:Napprox2}).
Nonetheless  if we extrapolate Eq.~(\ref{eqn:fitQ2}) to the true
Earth-Moon system where $r_{sB}/d=0.0046$ the correction from full screening $Q_2=-1$ remains only at the percent level.

\section{Discussion}
\label{sec:conclusion}

In this paper we studied the apparent violation of the equivalence
principle in a model with the Vainshtein mechanism where nonlinear
derivative interactions of the field act to screen the fifth force. We
considered the motion of two bodies that are separated by much less than
their individual Vainshtein radii such as the Earth-Moon system. In such
a system, the small body $B$ cannot be considered as a test body in the
large body $A$'s field $\phi_A$. The nonlinear equation for the field
allows for screening solutions where the Laplacian of $\phi_A$ is
screened within a region much larger in extent than the physical size of
body $B$. The primary effect is that the Laplacian $\nabla^2\phi_A$ is
fully screened. A crude estimate of the residual impact on the first
derivatives or forces would suggest a suppression of order
$(M_B/M_A)^{2/3}$ where $M_B/M_A$ is the mass ratio of the two-bodies.
Moreover, in the limit where the Vainshtein radii $r_{* A,B}$ are much
larger than the separation $d$ between the two bodies, we expect
relative screening effects to be independent of the crossover scale $r_c$.  

We confirmed these qualitative expectations by solving the joint two
body system numerically. Cast in terms of the deviation of the true
field from the superposition solution 
$\phi_{\Delta} = \phi - \phi_A -\phi_B$, in the interior of body $B$
there is nearly full screening 
$\nabla^2 \phi_{\Delta} \approx -\nabla^2 \phi_A$ in the limit that 
the size of body $B$ is much smaller than the separation between the
bodies.  On the other hand, the screening of forces from the large body
on the small body depend almost exclusively on the mass ratio
$M_B/M_A$. From numerical solutions, we found it is given by 
$Q_1(0) \approx -0.56 (M_B/M_A)^{0.6}$. 

The fifth force introduces an additional contribution to the anomalous
perihelion precession rate. In the DGP normal branch, the precession
rate was obtained by ignoring the nonsuperimposability and treating the
small body (e.g., Moon) as a test body. The precession rate (the angle
of perihelion advanve $\Delta\varphi_{\rm DGP}$ during one orbital
period $P$) is universal under this assumption and given by 
\begin{equation}
{\Delta \varphi_{\rm DGP} \over P} = {3 \over 8}{1 \over r_c} =  7.91 \left( {h \over H_0 r_c} \right) \mu{\rm arcsec}/{\rm yr}.
\end{equation}
This result needs to be revisited in the light of our finding. As
screening operates by replacing the large body's field with a linearized
version of itself, one might think that this affects the precession rate
qualitatively by changing the radial dependence of the force
law. However, the fractional effect is independent of the separation $d$
and the impact of nonsuperimposability comes from the field at the
position of the small body $B$, not at a fixed $r$. Then the only effect
of screening is to reduce the large body's force by the factor of $1
+Q_1$ and we have a proportional change in the precession, 
\begin{equation}
{\Delta \varphi_{\rm DGP} \over P} = {3 \over 8}{1 \over r_c} (1 + Q_1).
\end{equation}
Precession thus depends on the mass ratio of the bodies and is not universal.   Different mass bodies will precess at different rates leading to an apparent equivalence principle violation.   On the other hand, the mass ratio scaling of the equivalence principle violation implies that for typical systems the effect will be small.   For example, for
the mass ratio of the Earth-Moon system  $M_B/M_A = 1/80$, leading to a small deviation ($4\%$) from the universal precession rate.   Nonetheless, in principle the Vainshtein mechanism can be tested by precision tests of the perihelion precession of different mass objects.


\begin{acknowledgments}
We thank Alexander Belikov and Yudai Suwa for useful discussions. T.~H. was
supported by JSPS Grant-in-Aid for Young Scientists (B) No.~23740186,
 partially by JSPS Grant-in-Aid for Scientific Research (A) No.~21244033,
 and also by MEXT HPCI Strategic Program. W.~H. was  supported by
 the U.S.~Department of Energy Contract No.~DE-FG02-90ER-40560, the Kavli Institute
 for Cosmological Physics at the University of Chicago through Grants
 No.~NSF PHY-0114422 and No.~NSF PHY-0551142 and an endowment from the Kavli 
 Foundation and its founder Fred Kavli, and the David and Lucile Packard
 Foundation. K.~K. was supported by STFC Grant No.~ST/H002774/1, the ERC and the
 Leverhulme trust. F.~S. was supported by the Gordon and Betty Moore
 Foundation at Caltech. W.~H., K.~K., and F.~S. thank the Aspen Center
 for Physics 
 where this project was initiated and the organizers of the Ringberg
 dark energy meeting where it was completed. K.~K. thanks the Kavli
 Institute for Cosmological Physics at the University of Chicago for its
 hospitality.
\end{acknowledgments}

\appendix
\section{Numerical Techniques}
\label{sec:setup}

\subsection{Cylindrical Coordinates}

The field equation for the two-body deviation from superposition $\phi_\Delta$ given by
Eq.~(\ref{eqn:phideltaeom}) has cylindrical symmetry along the $z$ axis separating
the bodies (see Fig.~\ref{fig:compdom})
and so
%
\begin{equation}
\begin{aligned}
&\left[
     1
   + \frac{2r_c^2}{3}\left(
         \frac{1}{r}\pfrac{\phi_{AB}}{r}
         + \ppfrac{\phi_{AB}}{z}
     \right)
\right] \ppfrac{\phi_\Delta}{r}
+\left[
     1
   + \frac{2r_c^2}{3}\left(
         \frac{1}{r}\pfrac{\phi_{AB}}{r}
         + \ppfrac{\phi_{AB}}{r}
     \right)
\right] \ppfrac{\phi_\Delta}{z}
\\
&+ \left[
  1
  + \frac{2r_c^2}{3}\left(
        \ppfrac{\phi_{AB}}{r}
      + \ppfrac{\phi_{AB}}{z}
  \right)
\right]\frac{1}{r}\pfrac{\phi_\Delta}{r}
-\frac{4r_c^2}{3}\ppmfrac{\phi_{AB}}{r}{z}\ppmfrac{\phi_\Delta}{r}{z}
+  \frac{1}{3}N[\phi_\Delta,\phi_\Delta] =
- \frac{2}{3}N[\phi_A,\phi_B],
\end{aligned} \label{eq:deleq}
\end{equation}
%
where $\phi_{AB}=\phi_A+\phi_B$ and we have used
%
\begin{equation}
 \nabla^2\phi_{a} = \ppfrac{\phi_a}{r} + \frac{1}{r}\pfrac{\phi_a}{r} + \ppfrac{\phi_a}{z},\\
\end{equation}\label{eq:Lapcyl}
%
and
%
\begin{equation}
\begin{aligned}
\frac{1}{r_c^2}N[\phi_a,\phi_b] &=
\frac{1}{r}\left(
    \pfrac{\phi_a}{r}\ppfrac{\phi_b}{r}
  + \ppfrac{\phi_a}{r}\pfrac{\phi_b}{r}
\right)
-2\ppmfrac{\phi_a}{r}{z}\ppmfrac{\phi_b}{r}{z}\\
&\quad +\left(
  \ppfrac{\phi_a}{r} + \frac{1}{r}\pfrac{\phi_a}{r}
\right)\ppfrac{\phi_b}{z}
+\ppfrac{\phi_a}{z}\left(
  \ppfrac{\phi_b}{r} + \frac{1}{r}\pfrac{\phi_b}{r}
\right),
\end{aligned}\label{eq:Ncyl}
\end{equation}
%
where $\phi_a,\phi_b \in \{\phi_A,\phi_B,\phi_{AB},\phi_\Delta \}$.
We treat $r_{gA}/d$,  $M_B/M_A$, $r_{sA}/d$, $r_{sB}/d$ and $r_c/d$ as the model
parameters as mentioned in Sec.~\ref{subsec:modelparams} and so our fundamental unit of
length is the separation $d$.

In our numerical computations, we impose the regularity condition
at $r=0$, given as $\partial\phi/\partial r=0$. As for the other
boundaries, we assume that superposition holds;   namely, we set $\phi_\Delta=0$ at the exterior boundaries. We
study the effects of the finite computational domain on the
numerical solution in Sec.~\ref{appsec:tests}.

\subsection{Nonlinear coordinates}
\label{subsec:coordinate}

To justify the assumption, $\phi_\Delta=0$ at the boundaries,
we have to use a box size $L \gg r_{*A}$.
However, we also want $r_{*B} \gg d$ and so it is quite difficult to simulate these conditions in the $(r,z)$ coordinate choice.

Since high resolution is only required near the bodies, we can extend the dynamic range with the help of a nonlinear transformation of the radial and axial variables
%
\begin{equation}
 r = \rho + \frac{\alpha}{3}\rho^3, \quad z=\zeta + \frac{\alpha}{3}\zeta^3,
 \label{eq:trans}
\end{equation}
%
where $\alpha$ is a constant controlling the growth of the grid spacing at large
distance. According to these equations, the spatial intervals in the
original coordinate are given by
%
\begin{eqnarray}
\frac{\Delta r}{\Delta \rho} &\equiv &A^{-1}(\rho) =  (1+\alpha\rho^2) ,\nonumber\\
\frac{\Delta z}{\Delta \zeta} &\equiv &B^{-1}(\zeta) =  (1+\alpha\zeta^2),
\end{eqnarray}
%
becoming large at large distance for fixed $\Delta\rho$ and
$\Delta\zeta$ relative to the finer resolution at the origin.
If we take $\alpha=0.1$, $\rho=20d$ corresponds to $r\approx 286.7d$.  Since
$r_{*A}/d=58.9$, this choice  satisfies our requirement, $r_{*A} \ll L$, and we take
it as the fiducial size of the computational domain (see Table \ref{tab:fidparam}).

In these rescaled coordinates, Eqs.~(\ref{eq:Lapcyl}) and (\ref{eq:Ncyl}) are
further transformed according to
%
\begin{alignat}{5}
 \pfrac{\phi}{r} =A\pfrac{\phi}{\rho}, &\quad
 &\ppfrac{\phi}{r} =A^2\ppfrac{\phi}{\rho}+AA'\pfrac{\phi}{\rho},\\
 \pfrac{\phi}{z} =B\pfrac{\phi}{\zeta}, &\quad
 &\ppfrac{\phi}{z} =B^2\ppfrac{\phi}{\zeta}+BB'\pfrac{\phi}{\zeta}, &\quad
 &\ppmfrac{\phi}{r}{z} =AB\ppmfrac{\phi}{\rho}{\zeta}.
\end{alignat}
%

\subsection{Discretization}

We discretize the computational domain as $\rho_i = (i+1/2)h$
and $\zeta_j = (j-N)h$ with $i=0,1,\ldots,N$ and $j=0,1,\ldots,2N$
where $h$ is the spatial interval.  Note that since
%
\begin{equation}
L = (N h) + \frac{\alpha}{3} (N h)^3,
\end{equation}
%
the fiducial parameter choices of Table~\ref{tab:fidparam} are achieved with
$\alpha=0.1$ and $N=1600$.

The derivatives with respect to $\rho$ and $\zeta$ are approximated by
the central finite differences on the grid as
%
\begin{equation}
\begin{aligned}
 \pfrac{\phi}{\rho}
   &\approx \frac{\phi_{i+1,j}-\phi_{i-1,j}}{2h},
 &\ppmfrac{\phi}{\rho}{\zeta}
   &\approx \frac{\phi_{i+1,j+1}-\phi_{i-1,j+1}-\phi_{i+1,j-1}+\phi_{i-1,j-1}}{4h^2}, \\
 \ppfrac{\phi}{\rho}
   &\approx \frac{\phi_{i+1,j}-2\phi_{i,j}+\phi_{i-1,j}}{h^2},
 &\ppfrac{\phi}{\zeta}
   &\approx \frac{\phi_{i,j+1}-2\phi_{i,j}+\phi_{i,j-1}}{h^2},
\end{aligned}\label{eq:fd}
\end{equation}
%
where we abbreviated $\phi_{i,j}=\phi(\rho_i,\zeta_j)$.
The reason why we shift the $\rho$ coordinate by half a spatial interval $h$ is
to be able to easily impose the Neumann boundary condition  at $\rho=0$.
Considering an auxiliary point at $i=-1$, the boundary condition reads
%
\begin{equation}
 \left.\frac{\partial\phi}{\partial\rho}\right|_{\rho=0} = 0
\quad \Longrightarrow \quad
 \frac{\phi_{0,j}-\phi_{-1,j}}{h} = 0,
\quad \therefore \phi_{-1,j} = \phi_{0,j}.
\end{equation}
%
Thus the boundary condition for the value of the auxiliary point enforces the vanishing of the derivative without modifying the finite difference scheme.
For example, the {radial term in the Laplacian} evaluated at
$i=0$ becomes
%
\begin{equation}
\begin{aligned}
\left.\left\{
  A^2\ppfrac{\phi}{\rho}
+ \left(AA'+\frac{A}{r}\right)\pfrac{\phi}{\rho}
\right\}\right|_{\rho=\rho_0}
&\approx
A^2\frac{\phi_{1,j}-2\phi_{0,j}+\phi_{-1,j}}{h^2}
+ \left(AA'+\frac{A}{r}\right)
\frac{\phi_{1,j}-\phi_{-1,j}}{2h} \\
&=
A^2\frac{\phi_{1,j}-\phi_{0,j}}{h^2}
+ \left(AA'+\frac{A}{r}\right)
\frac{\phi_{1,j}-\phi_{0,j}}{2h}.
\end{aligned}
\end{equation}
%
Equation (\ref{eq:deleq}) written in $(\rho,\zeta)$
coordinate can then be recast as a matrix equation, schematically given as
%
\begin{equation}
 ({\rm linear\,terms\,of}\,\phi_\Delta)
 = N[\phi_\Delta,\phi_\Delta]
   + N[\phi_A,\phi_B]
\;\;\longrightarrow\;\;
 A\xx = \widetilde{\bb} \equiv \mathbf{f}(\xx) + \bb, \label{eq:diffeq1}
\end{equation}
%
where the matrix $A$ represents the discrete linear operators on $\phi$
given by the combination of Eq.~(\ref{eq:fd}), being an
asymmetric nine-band sparse $(N+1)(2N+1)\times (N+1)(2N+1)$ matrix, and
${\bf b}$ and ${\bf f}$ are vectors containing
$N[\phi_A,\phi_B]$ and $[\phi_\Delta,\phi_\Delta]$
evaluated at $(\rho_i,\zeta_j)$.

\subsection{Iteration scheme}

To solve the set of nonlinear equations given in Eq.~(\ref{eq:diffeq1}),
we use a combination of preconditioned conjugate gradient
squared (CGS) and successive over-relaxation (SOR) methods according to
Ref.~\cite{Shibata:1997nc}.
Let the $n$th iterated solution be $\xx^{(n)}$.   We evaluate the
right-hand side of Eq.~(\ref{eq:diffeq1}), $\widetilde{\bb}^{(n)} = \bb + \mathbf{f}(\xx^{(n)})$.
Then we solve the linear equation, $A\xx^{*} = \widetilde{\bb}^{(n)}$, by
CGS. Finally, we update the solution $\xx^{(n+1)}$ by SOR as
%
\begin{equation}
 \xx^{(n+1)} = \omega \xx^{*} + (1-\omega)\xx^{(n)}, \label{eq:iteration}
\end{equation}
%
where we set $\omega=0.5\sim 1$ to make the iterative solution
converge.

To achieve the fast convergence of CGS, we precondition the matrix $A$ with
modified incomplete LU decomposition, referred to as MILUCGS in
the literature. We decompose $A$ such that $A=LDU+R$, where $L$, $U$ are
just the copies of lower and upper triangular part of $A$, and $D$ is a
diagonal matrix, which are given by
%
\begin{equation}
 L_{ij} =
   \begin{cases}
  0 & i<j \\
  d_i^{-1} & i=j \\
  A_{ij} & i>j
   \end{cases}, \quad
 U_{ij} =
   \begin{cases}
  A_{ij} & i<j \\
  d_i^{-1} & i=j \\
  0 & i>j
   \end{cases}, \quad
 D_{ij} = d_i\delta_{ij}.
\end{equation}
%
The diagonal element $d_i$ is recursively calculated by
%
\begin{equation}
 \begin{aligned}
 d_i^{-1} &= A_{ii}
     - A_{i,i-N}d_{i-N}(A_{i-N,i}+A_{i-N,i-N+1})\notag\\
&\qquad\quad- A_{i,i-1}d_{i-1}(A_{i-1,i}+A_{i-1,i+N-1}).\\
 \end{aligned}
\end{equation}
%
Multiplying $(LDU)^{-1}$ in both sides in Eq.~(\ref{eq:diffeq1}),
we obtain a new operator matrix $A'=(LDU)^{-1}A=I+(LDU)^{-1}R$
where we formally separated as $A=LDU+R$. The resultant matrix $A'$
becomes close to a unit matrix in the sense that the weight of
the nonzero components of $A'$ becomes significant near the diagonal.  In
other words, the condition number $\kappa(A')=||A'^{-1}||/||A'||$
becomes small. This treatment is frequently used to accelerate the
convergence of CGS.

We stop the SOR iteration in Eq.~(\ref{eq:iteration}) if we achieve
%
\begin{equation}
  \frac{||\widetilde{\bb}^{(n+1)} - A\xx^{(n+1)}||}{||\widetilde{\bb}^{(n+1)}||} < \epsilon_{\rm SOR},
  \label{eq:error}
\end{equation}
%
where $||\cdots||$ represents the standard 2-norm, and we set $\epsilon_{\rm SOR}=10^{-8}$.

\subsection{Convergence tests}
\label{appsec:tests}

\begin{figure}[!ht]
 \centering{
   \includegraphics[width=8.5cm]{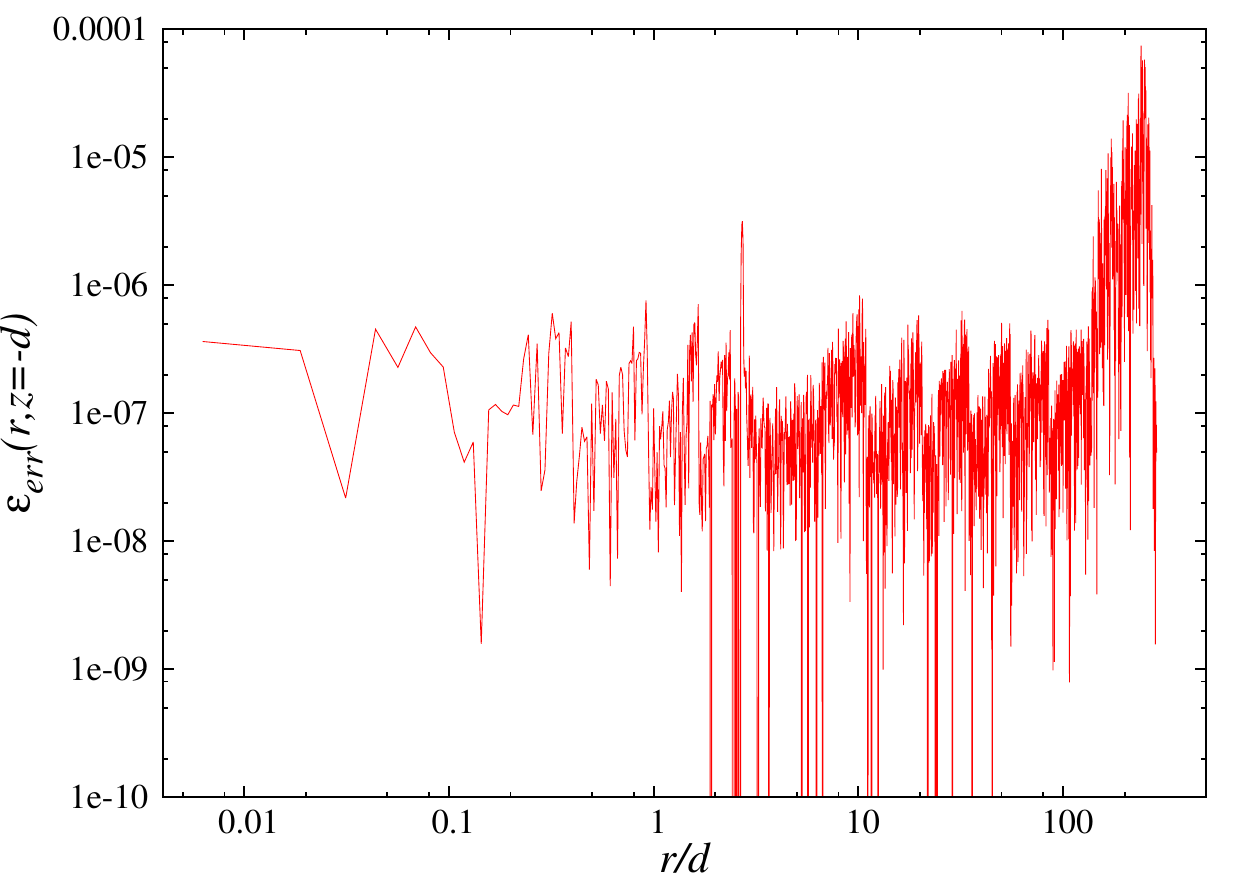}
  }
  \caption{
  The fractional accuracy of the numerical results with the fiducial choice of
  parameters along $z=-d$.
  }
  \label{fig:T37}
\end{figure}

First, we show the accuracy of our numerical results when we take the
fiducial choice of parameters. In Fig.~\ref{fig:T37}, we plot
the ratio of the residuals of the field equation
(\ref{eqn:phideltaeom}) to its source term,
%
\begin{equation}
\epsilon_{\rm err}(r,z) \equiv
\frac{3 \nabla^2 \phi_\Delta + N[\phi_\Delta,\phi_\Delta]
+ 2 N[\phi_A+\phi_B,\phi_\Delta] + 2N [\phi_A,\phi_B]}{2N [\phi_A,\phi_B]}.
\end{equation}
%
Setting $\epsilon_{\rm SOR}=10^{-8}$, we find that the
local residuals along $z=-d$ are suppressed to less than $10^{-6}$ except in the
far-field regime where they are still below $10^{-4}$.
Recall that the boundary conditions at the edges of the box are set to enforce
superposition.   Given finite computational resources there is a trade-off between
increased box size and central resolution.

\begin{figure}[!ht]
  \centering{
    \includegraphics[width=8.5cm]{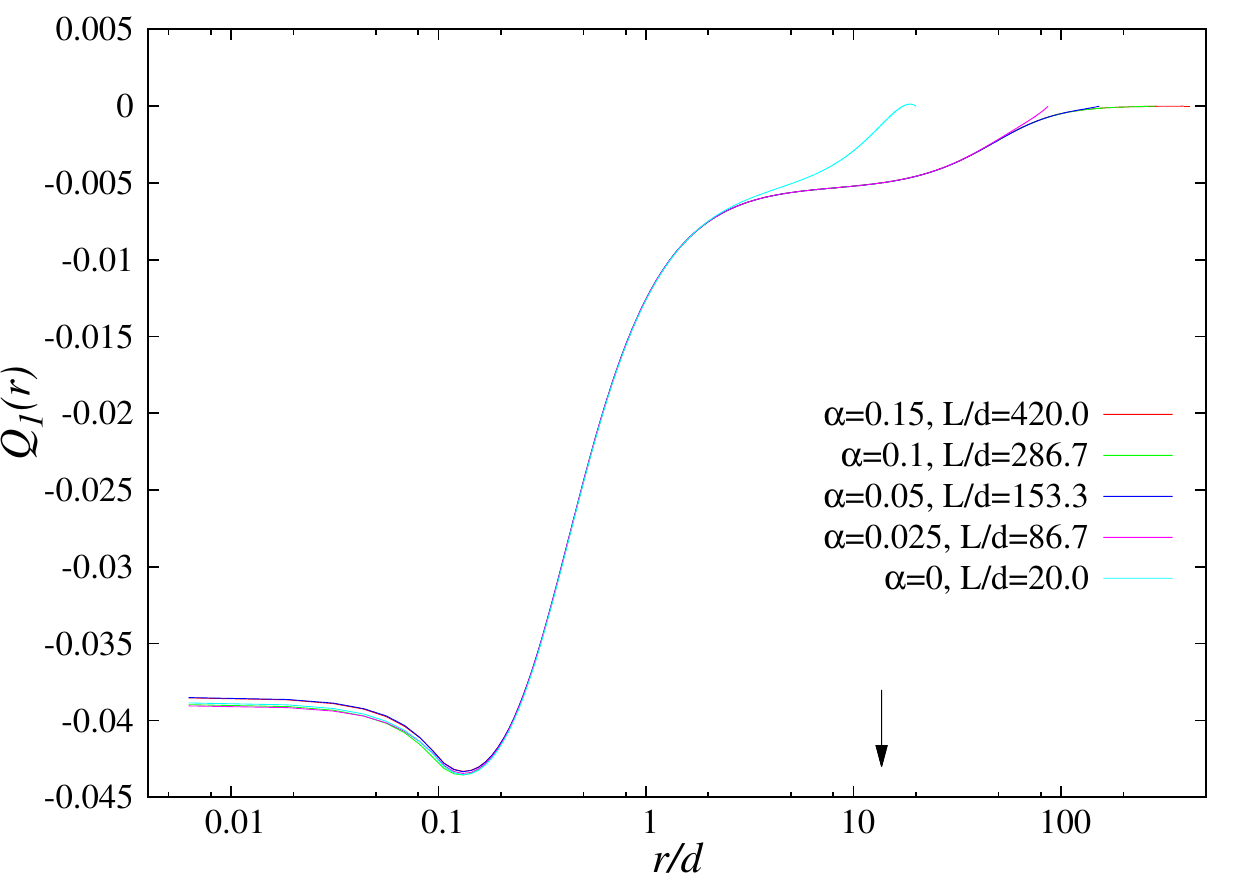} 
    \includegraphics[width=8.5cm]{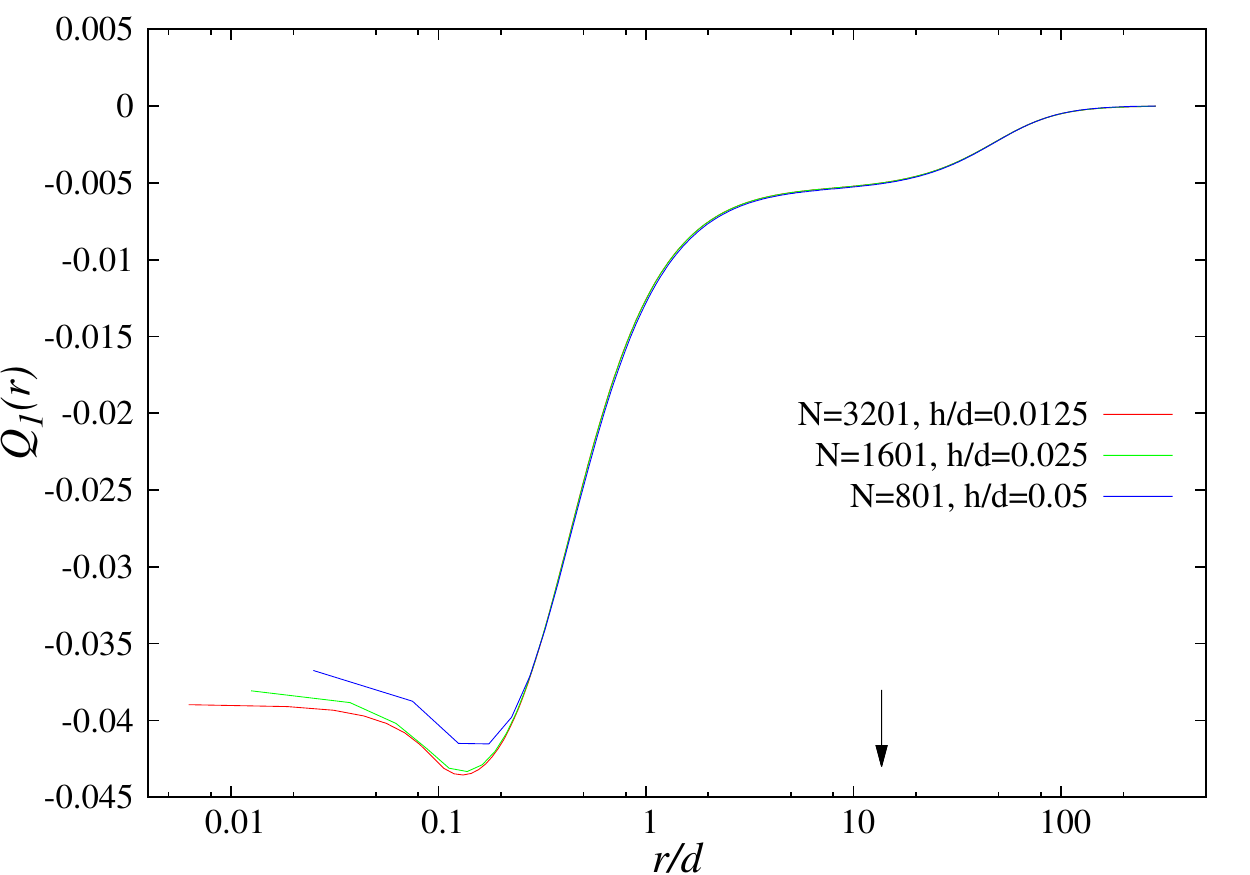} 
    \includegraphics[width=8.5cm]{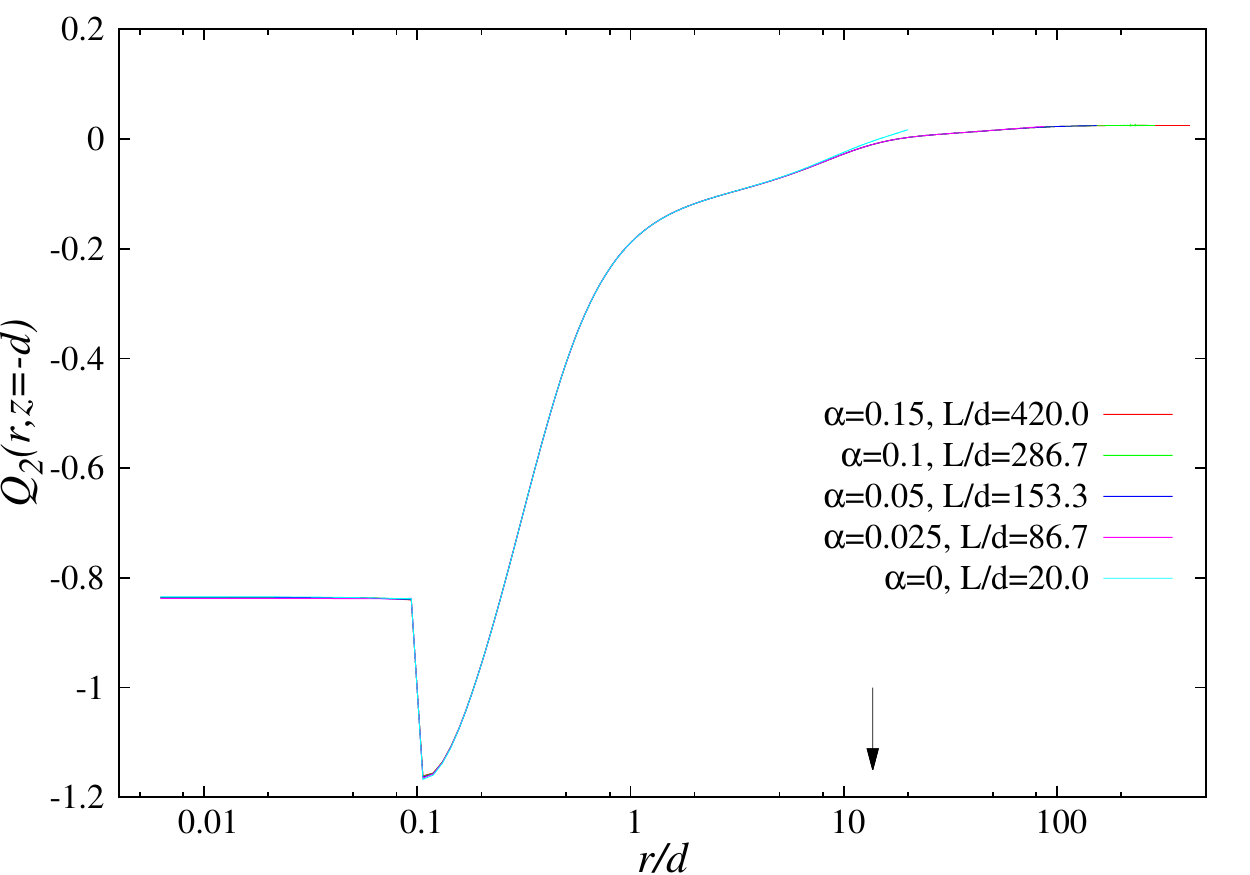} 
    \includegraphics[width=8.5cm]{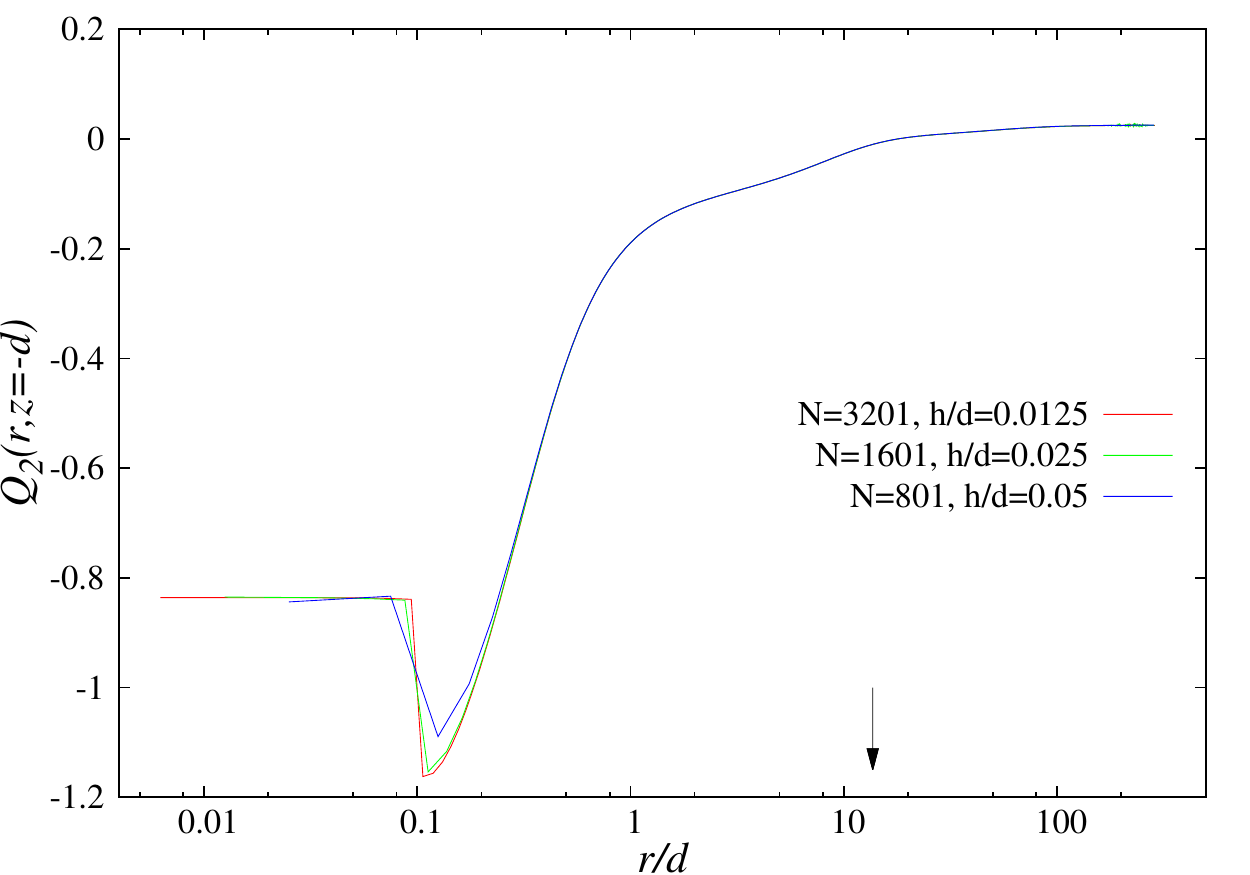} 
    }
  \caption{
    Box size and resolution dependence of screening statistics.  Left:
 changing $\alpha$ at fixed $N, h$ changes the box size $L$.   Results
 are independent of $L$ near the bodies if $L\gg r_{*B}$ (arrows).
 Right: changing $N$ at fixed $\alpha$ changes the resolution $h$.
 Results are independent of $h$ if $h \ll r_{sB}$.
  }
  \label{fig:T36}
\end{figure}

We therefore also test robustness of our results to the box size $L$
and resolution $h$.    In Fig.~\ref{fig:T36} (top), we vary $\alpha$
at fixed $h$ or $N$ thus changing $L$.   As long as $L \gg r_{*B}$
results near the body are independent of box size.  If $L> r_{*A}$ then
we regain the far-field behavior, justifying the use of superposition
boundary conditions.
In Fig.~\ref{fig:T36} (bottom), we study resolution $h$ by changing $N$
at fixed $\alpha$.   Interior to body $B$, a resolution of $h/r_{sB}\le
0.125$ is required to obtain converged results.

\bibliographystyle{apsrev}


\end{document}